\documentstyle[11pt,epsf]{article}
\begin{document}
\newcommand {\ee}{\end{equation}}
\newcommand {\bea}{\begin{eqnarray}}
\newcommand {\eea}{\end{eqnarray}}
\newcommand {\nn}{\nonumber \\}
\newcommand {\Tr}{{\rm Tr\,}}
\newcommand {\tr}{{\rm tr\,}}
\newcommand {\e}{{\rm e}}
\newcommand {\etal}{{\it et al.}}
\newcommand {\m}{\mu}
\newcommand {\n}{\nu}
\newcommand {\pl}{\partial}
\newcommand {\p} {\phi}
\newcommand {\vp}{\varphi}
\newcommand {\vpc}{\varphi_c}
\newcommand {\al}{\alpha}
\newcommand {\be}{\beta}
\newcommand {\ga}{\gamma}
\newcommand {\Ga}{\Gamma}
\newcommand {\x}{\xi}
\newcommand {\ka}{\kappa}
\newcommand {\la}{\lambda}
\newcommand {\La}{\Lambda}
\newcommand {\si}{\sigma}
\newcommand {\Si}{\Sigma}
\newcommand {\th}{\theta}
\newcommand {\Th}{\Theta}
\newcommand {\om}{\omega}
\newcommand {\Om}{\Omega}
\newcommand {\ep}{\epsilon}
\newcommand {\vep}{\varepsilon}
\newcommand {\na}{\nabla}
\newcommand {\del}  {\delta}
\newcommand {\Del}  {\Delta}
\newcommand {\mn}{{\mu\nu}}
\newcommand {\ls}   {{\lambda\sigma}}
\newcommand {\ab}   {{\alpha\beta}}
\newcommand {\half}{ {\frac{1}{2}} }
\newcommand {\third}{ {\frac{1}{3}} }
\newcommand {\fourth} {\frac{1}{4} }
\newcommand {\sixth} {\frac{1}{6} }
\newcommand {\sqg} {\sqrt{g}}
\newcommand {\fg}  {\sqrt[4]{g}}
\newcommand {\invfg}  {\frac{1}{\sqrt[4]{g}}}
\newcommand {\sqZ} {\sqrt{Z}}
\newcommand {\gbar}{\bar{g}}
\newcommand {\sqk} {\sqrt{\kappa}}
\newcommand {\sqt} {\sqrt{t}}
\newcommand {\reg} {\frac{1}{\epsilon}}
\newcommand {\fpisq} {(4\pi)^2}
\newcommand {\Lcal}{{\cal L}}
\newcommand {\Ocal}{{\cal O}}
\newcommand {\Dcal}{{\cal D}}
\newcommand {\Ncal}{{\cal N}}
\newcommand {\Mcal}{{\cal M}}
\newcommand {\scal}{{\cal s}}
\newcommand {\Dvec}{{\hat D}}   
\newcommand {\dvec}{{\vec d}}
\newcommand {\Evec}{{\vec E}}
\newcommand {\Hvec}{{\vec H}}
\newcommand {\Vvec}{{\vec V}}
\newcommand {\Btil}{{\tilde B}}
\newcommand {\ctil}{{\tilde c}}
\newcommand {\Ftil}{{\tilde F}}
\newcommand {\Stil}{{\tilde S}}
\newcommand {\Ztil}{{\tilde Z}}
\newcommand {\altil}{{\tilde \alpha}}
\newcommand {\betil}{{\tilde \beta}}
\newcommand {\latil}{{\tilde \lambda}}
\newcommand {\ptil}{{\tilde \phi}}
\newcommand {\Ptil}{{\tilde \Phi}}
\newcommand {\natil} {{\tilde \nabla}}
\newcommand {\ttil} {{\tilde t}}
\newcommand {\Rhat}{{\hat R}}
\newcommand {\Shat}{{\hat S}}
\newcommand {\shat}{{\hat s}}
\newcommand {\Dhat}{{\hat D}}   
\newcommand {\Vhat}{{\hat V}}   
\newcommand {\xhat}{{\hat x}}
\newcommand {\Zhat}{{\hat Z}}
\newcommand {\Gahat}{{\hat \Gamma}}
\newcommand {\nah} {{\hat \nabla}}
\newcommand {\gh}  {{\hat g}}
\newcommand {\labar}{{\bar \lambda}}
\newcommand {\cbar}{{\bar c}}
\newcommand {\bbar}{{\bar b}}
\newcommand {\Bbar}{{\bar B}}
\newcommand {\psibar}{{\bar \psi}}
\newcommand {\chibar}{{\bar \chi}}
\newcommand {\bfZ} {{\bf Z}}
\newcommand {\bfR} {{\bf R}}
\newcommand {\bbartil}{{\tilde {\bar b}}}
\newcommand  {\vz}{{v_0}}
\newcommand {\intfx} {{\int d^4x}}
\newcommand {\inttx} {{\int d^2x}}
\newcommand {\change} {\leftrightarrow}
\newcommand {\ra} {\rightarrow}
\newcommand {\larrow} {\leftarrow}
\newcommand {\ul}   {\underline}
\newcommand {\pr}   {{\quad .}}
\newcommand {\com}  {{\quad ,}}
\newcommand {\q}    {\quad}
\newcommand {\qq}   {\quad\quad}
\newcommand {\qqq}   {\quad\quad\quad}
\newcommand {\qqqq}   {\quad\quad\quad\quad}
\newcommand {\qqqqq}   {\quad\quad\quad\quad\quad}
\newcommand {\qqqqqq}   {\quad\quad\quad\quad\quad\quad}
\newcommand {\qqqqqqq}   {\quad\quad\quad\quad\quad\quad\quad}
\newcommand {\lb}    {\linebreak}
\newcommand {\nl}    {\newline}

\newcommand {\vs}[1]  { \vspace*{#1 cm} }

\newcommand {\MPL}  {Mod.Phys.Lett.}
\newcommand {\NP}   {Nucl.Phys.}
\newcommand {\PL}   {Phys.Lett.}
\newcommand {\PR}   {Phys.Rev.}
\newcommand {\PRL}   {Phys.Rev.Lett.}
\newcommand {\CMP}  {Commun.Math.Phys.}
\newcommand {\JMP}  {Jour.Math.Phys.}
\newcommand {\AP}   {Ann.of Phys.}
\newcommand {\PTP}  {Prog.Theor.Phys.}
\newcommand {\NC}   {Nuovo Cim.}
\newcommand {\CQG}  {Class.Quantum.Grav.}


\font\smallr=cmr5
\def\ocirc#1{#1^{^{{\hbox{\smallr\llap{o}}}}}}
\def\ogamma{\ocirc{\gamma}{}}
\def\oM{{\buildrel {\hbox{\smallr{o}}} \over M}}
\def\osigma{\ocirc{\sigma}{}}

\def\overleftrightarrow#1{\vbox{\ialign{##\crcr
 $\leftrightarrow$\crcr\noalign{\kern-1pt\nointerlineskip}
 $\hfil\displaystyle{#1}\hfil$\crcr}}}
\def\overnab{{\overleftrightarrow\nabslash}}

\def\va{{a}}
\def\vb{{b}}
\def\vc{{c}}
\def\tilpsi{{\tilde\psi}}
\def\tbpsi{{\tilde{\bar\psi}}}

\def\Dslash{{}\hbox{\hskip2pt\vtop
 {\baselineskip23pt\hbox{}\vskip-24pt\hbox{/}}
 \hskip-11.5pt $D$}}
\def\nabslash{{}\hbox{\hskip2pt\vtop
 {\baselineskip23pt\hbox{}\vskip-24pt\hbox{/}}
 \hskip-11.5pt $\nabla$}}
\def\xislash{{}\hbox{\hskip2pt\vtop
 {\baselineskip23pt\hbox{}\vskip-24pt\hbox{/}}
 \hskip-11.5pt $\xi$}}
\def\leftnabla{{\overleftarrow\nabla}}

\def\delL{{\delta_{LL}}}
\def\delG{{\delta_{G}}}
\def\delc{{\delta_{cov}}}

\newcommand {\sqxx}  {\sqrt {x^2+1}}   
\newcommand {\gago}  {\gamma_5}
\newcommand {\Ktil}  {{\tilde K}}
\newcommand {\Ltil}  {{\tilde L}}
\newcommand {\Qtil}  {{\tilde Q}}
\newcommand {\Rtil}  {{\tilde R}}
\newcommand {\Kbar}  {{\bar K}}
\newcommand {\Lbar}  {{\bar L}}
\newcommand {\Qbar}  {{\bar Q}}
\newcommand {\Pp}  {P_+}
\newcommand {\Pm}  {P_-}
\newcommand {\GfMp}  {G^{5M}_+}
\newcommand {\GfMpm}  {G^{5M'}_-}
\newcommand {\GfMm}  {G^{5M}_-}
\newcommand {\Omp}  {\Omega_+}    
\newcommand {\Omm}  {\Omega_-}
\def\Aslash{{}\hbox{\hskip2pt\vtop
 {\baselineskip23pt\hbox{}\vskip-24pt\hbox{/}}
 \hskip-11.5pt $A$}}
\def\Rslash{{}\hbox{\hskip2pt\vtop
 {\baselineskip23pt\hbox{}\vskip-24pt\hbox{/}}
 \hskip-11.5pt $R$}}
\def\kslash{
{}\hbox       {\hskip2pt\vtop
                   {\baselineskip23pt\hbox{}\vskip-24pt\hbox{/}}
               \hskip-8.5pt $k$}
           }    
\def\qslash{
{}\hbox       {\hskip2pt\vtop
                   {\baselineskip23pt\hbox{}\vskip-24pt\hbox{/}}
               \hskip-8.5pt $q$}
           }    
\def\dslash{
{}\hbox       {\hskip2pt\vtop
                   {\baselineskip23pt\hbox{}\vskip-24pt\hbox{/}}
               \hskip-8.5pt $\partial$}
           }    
\def\dbslash{{}\hbox{\hskip2pt\vtop
 {\baselineskip23pt\hbox{}\vskip-24pt\hbox{$\backslash$}}
 \hskip-11.5pt $\partial$}}
\def\Kbslash{{}\hbox{\hskip2pt\vtop
 {\baselineskip23pt\hbox{}\vskip-24pt\hbox{$\backslash$}}
 \hskip-11.5pt $K$}}
\def\Ktilbslash{{}\hbox{\hskip2pt\vtop
 {\baselineskip23pt\hbox{}\vskip-24pt\hbox{$\backslash$}}
 \hskip-11.5pt ${\tilde K}$}}
\def\Ltilbslash{{}\hbox{\hskip2pt\vtop
 {\baselineskip23pt\hbox{}\vskip-24pt\hbox{$\backslash$}}
 \hskip-11.5pt ${\tilde L}$}}
\def\Qtilbslash{{}\hbox{\hskip2pt\vtop
 {\baselineskip23pt\hbox{}\vskip-24pt\hbox{$\backslash$}}
 \hskip-11.5pt ${\tilde Q}$}}
\def\Rtilbslash{{}\hbox{\hskip2pt\vtop
 {\baselineskip23pt\hbox{}\vskip-24pt\hbox{$\backslash$}}
 \hskip-11.5pt ${\tilde R}$}}
\def\Kbarbslash{{}\hbox{\hskip2pt\vtop
 {\baselineskip23pt\hbox{}\vskip-24pt\hbox{$\backslash$}}
 \hskip-11.5pt ${\bar K}$}}
\def\Lbarbslash{{}\hbox{\hskip2pt\vtop
 {\baselineskip23pt\hbox{}\vskip-24pt\hbox{$\backslash$}}
 \hskip-11.5pt ${\bar L}$}}
\def\Rbarbslash{{}\hbox{\hskip2pt\vtop
 {\baselineskip23pt\hbox{}\vskip-24pt\hbox{$\backslash$}}
 \hskip-11.5pt ${\bar R}$}}
\def\Qbarbslash{{}\hbox{\hskip2pt\vtop
 {\baselineskip23pt\hbox{}\vskip-24pt\hbox{$\backslash$}}
 \hskip-11.5pt ${\bar Q}$}}
\def\Acalbslash{{}\hbox{\hskip2pt\vtop
 {\baselineskip23pt\hbox{}\vskip-24pt\hbox{$\backslash$}}
 \hskip-11.5pt ${\cal A}$}}

\begin{flushright}
September 17, 2001\\
hep-th/0008245 \\
US-00-07
\end{flushright}

\vspace{0.5cm}

\begin{center}
{\Large\bf 
Wall and Anti-Wall in\\
the Randall-Sundrum Model \\
and \\
A New Infrared Regularization}

\vspace{1.5cm}
{
\large Shoichi ICHINOSE
          \footnote{
E-mail address:\ ichinose@u-shizuoka-ken.ac.jp
                  }
}
\vspace{1cm}

{\large 
Laboratory of Physics, \\
School of Food and Nutritional Sciences, \\
University of Shizuoka,
Yada 52-1, Shizuoka 422-8526, Japan          
}

\end{center}
\vfill

{\large Abstract}\nl
An approach to find the field equation solution of
the Randall-Sundrum model with the $S^1/Z_2$ extra
axis is presented. We closely examine the infrared
singularity. The vacuum is set by the 5 dimensional Higgs field.
Both the domain-wall and the anti-domain-wall naturally
appear, at the {\it ends} of the extra compact axis, 
by taking a {\it new infrared regularization}.
The stability is guaranteed from the outset by
the kink boundary condition. 
A {\it continuous} (infrared-)regularized solution, which
is a truncated {\it Fourier series} of a {\it discontinuous}
solution, is utilized.The ultraviolet-infrared relation
appears in the regularized solution.
\vspace{0.5cm}

PACS NO:\ 04.25.-g,\ 04.50.+h,\ 11.10.Kk,\ 11.25.Mj,\ 
11.27.+d,\ 12.10.Kt\nl


\section{Introduction}
As an approach to explain the mass hierarchy problem,
the Randall-Sundrum (RS) model\cite{RS9905,RS9906} have
been taking people's attention both from the \nl
phenomenology\cite{Hewett00}
and from the theory\cite{KL0001,GL0003}. 
The model has, in fact, some advantages compared with
other approaches such as the Kaluza-Klein compactification
\cite{ADD98,AADD98} and the (standard) renormalization
group approach      
. The most characteristic
point is its exponential damping factor (warp factor) which
could have the possibility of 
naturally explaining the broadly-spreading mass hierarchy
ranging from the cosmological constant ($10^{-41}$GeV),
through the weak physics ($10^{2}$GeV), to the Planck
mass ($10^{19}$GeV). Furthermore the recent progress
in the AdS/CFT correspondence\cite{Mald98,Witt98,dBVV9912}
indicates the RS-model solution,
which is a {\it classical} solution in the 5 dim
AdS space-time, could be regarded as 
the renormalization trajectory in the 4 dim {\it quantum} solution. 

We point out, however, an incomplete aspect in most
approaches so far. They assume the $\del$-function
or $\th$-function distribution from the outset
as a form of the classical
solution in order to make a ("infinitely-thin") wall configuration. 
Indeed it gives an easy "tool"
to analyse the model in some limitted situation. It is, however,
obscure from the standpoint of the soliton (kink) physics
and does miss the important role of the "thickness"
in the regularization standpoint.
The configuration, considered in the RS-model, generally
has a domain wall structure with some {\it finite}
thickness which is determined by
the vacua in the asymptotic regions (or the boundary conditions)
and some parameters in the system. 
In some limitted configuration, the thickness approaches
zero and 
the  $\del$-function (or $\th$-function) appears as a 
well-regularized object.In such a way, 
we can understand the real meaning of the limit
from the vacuum structure or the system parameters.
This looks very important especially 
to understand the problem of the cosmological constant,
which is the vacuum energy of the space-time.
Needless to say, 
the system configuration should be derived by solving
the field equation in a proper way. 
In some reference\cite{CEHS0001}, the thickness was introduced
just by smearing the assumed $\del$-function. Such approach
loses the real role of the thickness. 

Motivated by the above things, a solution of
the RS-model, for the one-wall case, has been presented
\cite{SI00apr,ICHEP2000}. 
\begin{figure}
\centerline{\epsfysize=4cm\epsfbox{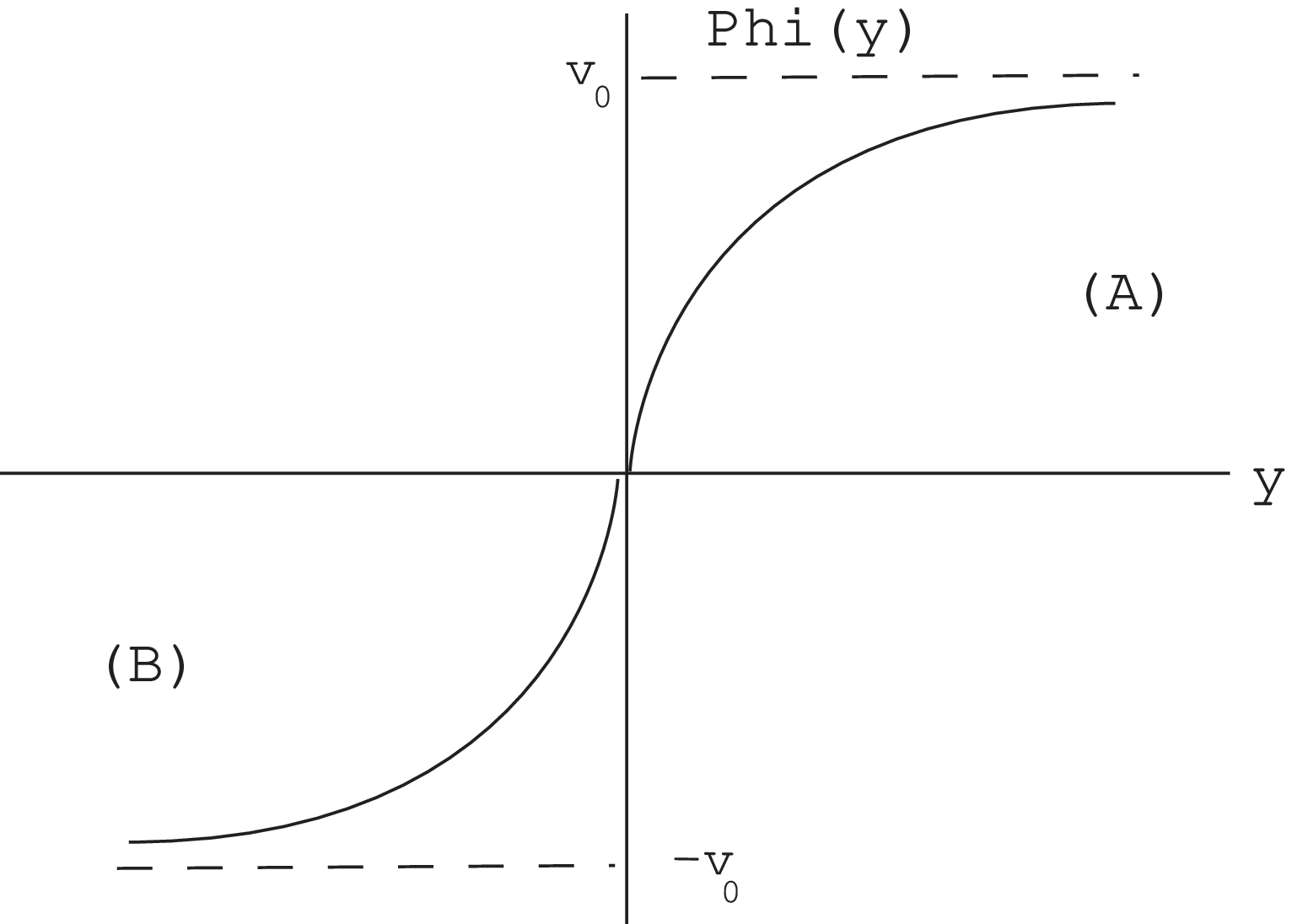}}
   \begin{center}
Fig.1\ The scalar field for one-wall configuration
\cite{SI00apr}. $y\in \bfR=(-\infty,\infty)$.
   \end{center}
\end{figure}
The points are\ 
1)\ the wall configuration is obtained as a kink solution of
the classical field equation of the 5 dim AdS gravity;\ 
2)\ the $\del$-function limit is specified by some parameters;\ 
3)\ the vacua (asymptotic states), which are 
necessary to specify the kink, are introduced
by the 5 dim bulk Higgs potential;\ 
4)\ the stability of the solution is guaranteed by the boundary
condition. 
The obtained solution correctly gives the basic physical
outputs such as the 4D Planck mass
($M_{pl}\sim \sqrt{M^3/k}$) and the 4D cosmological constant
($\La_{4d}\sim -M^3k$) in terms of the 5D Planck mass ($M$)
and the thickness parameter ($k$). These properties remain
valid in the present case because we 
modify (with much care) only the infrared boundary condition.
The configuration of a wall and an anti-wall is taken
in the original work\cite{RS9905}, but it is now considered
unstable. To cure it, one approach is to take into account
the radion field\cite{GW9907b}. Here we point out another
possibility of creating the wall-anti-wall configuration
by taking a new infrared treatment which mimics the lattice
situation.

It is well-known that the massless chiral fermion appears
as a zero mode bound to the domain wall. 
Now we recall the similar situation takes place in the lattice
field theory using the 5D lattice\cite{Kap92,Jan92}.
In that case, besides the fact that the formulation is discrete, 
some essential differences are there. That is, the extra axis $y$
is regularized to be {\it finite} $-L\leq y\leq +L$ and 
the boundary condtion for the extra axis is taken to be {\it periodic}:
\ $y\ra y+2L$. In this case a wall appears at $y=0$ and an anti-wall
appears at $y=L$. For every zero mode at $y=0$, there is a zero mode
of the opposite chirality at $y=L$. Every mode is {\it chirally paired}, 
which shows the {\it vector}-like nature of the 5D theory. 
The chiral anomaly in the 4D theory is now understood as
the flow of the current through the fifth
space\cite{CH85}. In the lattice numerical simulation, 
this configuration is basically taken, with some improvements
\cite{Sha93,CH94,Vra98}, and several physical quantities of QCD
such as the pion mass are numerically calculated\cite{Col0007}. 
In order to realize the similar situation in the Randall-Sundrum model,
we need
the wall-anti-wall configuration. We present a way to make
the configuration from the one-wall solution of Ref.\cite{SI00apr}.

A focus here is to clarify some controversial point,
that is, whether 
$S^1/Z_2$ compactification of 
the kink configuration(Fig.1) is compatible with
the wall-anti-wall configuration.
On the one hand, $S^1$ property requires the solution
to be periodic with some finite periodicity in
an extra axis.
On the other, naive expectation implies 
its behavior, in an asymptotic region (A), 
does {\it not continuously} connect with that in (B) of their
adjacent period, as far as (A) and (B) are {\it different vacua}
which is required for the soliton (kink) configuration. 
Clearly this is related to the boundary and stability problems.
The stability is guaranteed by the kink property:\ 
the two vacua (A) and (B) are related by the discrete
(discontinuous) symmetry, $\Phi\change -\Phi$. To solve
these problems,  
some close {\it infrared} treatment is necessary. 
Note that physical behaviors in both regions (A) and (B)
are the same. (Both have the same 5D scalar Riemann curvature.)
Further note that we do {\it not} consider a new solution
of the kink-anti-kink type. 
We will first choose a correct coordinate where 
the wall-anti-wall configuration should appear and then
 {\it change} the infrared
boundary condition for the one-wall solution previously
obtained. We will show
the new boundary condition gives us the other (anti-) wall. 
The present claim is that
the (stable) wall-anti-wall solution exists by taking
the correct coordinate and 
the {\it new infrared regularization} proposed here.

The present model of 5D gravity-scalar system
is introduced 
in Sec.2. In Sec.3 we change the extra coordinate from
the infinite one $y\in \bfR=(-\infty,\infty)$ to a compact
one $z\in (-\half r_c, +\half r_c)$ in order to obtain
the "size" of the extra space $r_c$. In Sec.4, by imposing
the periodic boundary condition, we extend the coordinate
region of $z$ to $\bfR$. The new infrared regularization is
explained in Sec.5, where the Fourier expansion of (continuous
and discontinuous) periodic functions is exploited. 
Truncation of the infinitely expanded terms to the finite ones 
is the key of the present regularization. In Sec.6, we present
the numerical solutions of the present model, and using this result, 
we fix all Fourier expansion coefficients numerically. We conclude
and discuss in Sec.7. Some appendices are in order to supplement
the text. In App.A, a compact coordinate, which is different from
the one taken in  the text, is examined. Some advantageous points
are noticed. Numerical results of Sec.6 are explained in App.B. 
They consist of two standard method of the numerical calculus:\ 
the Runge-Kutta method and the Least Square method. In App.C, some
simple function, which imitates the solution of the present model,
is examined in order to clarify the characteristic properties
of the Fourier expansion coefficients.

%
%
\section{Model Set-Up}
We take the following 5D gravitational theory with 5D Higgs
potential.
\begin{eqnarray}
S[G_{AB},\Phi]=\int d^5X\sqrt{-G} (-\half M^3\Rhat
-\half G^{AB}\pl_A\Phi\pl_B\Phi-V(\Phi))\com\nn
V(\Phi)=\frac{\la}{4}(\Phi^2-{v_0}^2)^2+\La\com
\label{int1}
\end{eqnarray}
where $X^A (A=0,1,2,3,4)$ is the 5D coordinates and we also use
the notation $(X^A)\equiv (x^\m,y), \m=0,1,2,3$. The coordinate 
$X^4=y$ is the extra axis which is taken to be a space coordinate.
$\Phi$ is a 5D scalar field, $G=\det G_{AB}$, $\Rhat$ is the
5D Riemannian scalar curvature. $M(>0)$ is the 5D Planck mass
and is regarded as the {\it fundamental scale} of this dimensional reduction
scenario. $V(\Phi)$ is the Higgs potential and serves for preparing
the (classical) vacuum in 5D world. 
The three parameters $\la,\vz$ and $\La$ in $V(\Phi)$ are called here
{\it vacuum parameters}. 
$\la(>0)$ is a coupling, $v_0(>0)$ 
is the Higgs field vacuum expectation value, 
and $\La$ is the 5D cosmological constant. See Fig.2.
\begin{figure}
\centerline{\epsfysize=4cm\epsfbox{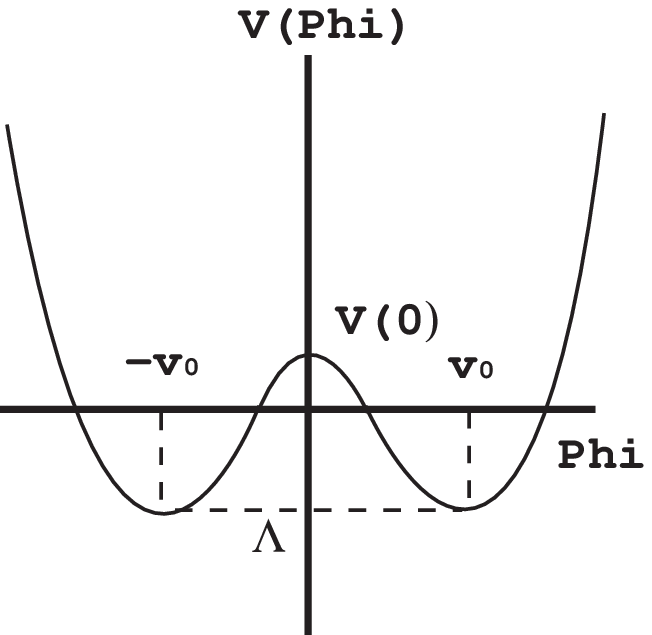}}
   \begin{center}
Fig.2\ The Higgs Potential $V(\Phi)$,
(\ref{int1}). Horizontal axis: $\Phi$. From (\ref{S.12}),
$V(0)=\la\vz^4/4+\La>0,\ V(\vz)=\La<0$.
   \end{center}
\end{figure}
It is later shown that the sign of $\La$ must be negative for
the present domain wall configuration.
The Einstein equation is given by
\begin{eqnarray}
M^3(\Rhat_{MN}-\half G_{MN}\Rhat )=-\pl_M\Phi\,\pl_N\Phi
+G_{MN}(\half G^{KL}\pl_K\Phi\,\pl_L\Phi+V(\Phi))\com\nn
\na^2\Phi=\frac{\del V}{\del \Phi}\pr
\label{int2}
\end{eqnarray}
Following Callan and Harvey\cite{CH85},
we consider the case that $\Phi$ depends only on 
the extra coordinate $y$:\ $\Phi=\Phi(y)$.
Because $M$-dependence can be absorbed by a simple
scaling 
($
\Phi=M^{3/2}{\tilde \Phi}, 
v_0=M^{3/2}{\tilde v_0}, \la=M^{-1}{\tilde \la},
\La=M^{5}{\tilde \La}, X^A=M^{-1}{\tilde X^A},
$)
we may, 
for simplicity, take
\begin{eqnarray}
M=1
\pr\label{int3}
\end{eqnarray}
We explicitly write $M$ only when it is necessary.
%
%
\section{Infinite Extra Axis and Its Compactification}
We start with the following 5D metric\cite{RS9906,SI00apr}.
\begin{eqnarray}
{ds}^2=\e^{-2\si(y)}\eta_\mn dx^\m dx^\n+{dy}^2\com\q
y\in {\bf R}=(-\infty,+\infty)
\label{R.1}
\end{eqnarray}
where $\eta_\mn=\mbox{diag}(-1,1,1,1)$. In this choice, the 4D Poincar{\' e}
invariance is preserved. The Weyl factor $\e^{-2\si(y)}$ is called
"warp factor" and is determined by the 5D Einstein equation.
The extra axis taken here is an infinite real line 
${\bf R}=(-\infty,+\infty)$. The coordinates 
$(X^A)=(x^\m,y)$ give one wall configuration by taking
the boundary condition:\ 
$\Phi(y)\ra\pm v_0\ ,\ y\ra\pm\infty$
and there exists a family of exact solution
\cite{SI00apr}.
Now let us move from the $y$-coordinate
to another one $z$(See Fig.3).
\begin{eqnarray}
\frac{Mz}{Mr_c}=\frac{z}{r_c}=\half \tanh ({My})\com\q
\frac{z}{r_c}\in (-\half,\half)
\com
\label{R.2}
\end{eqnarray}
where a {\it free} parameter
$r_c$ is introduced as the {\it compactification size}.
(Another compactification is examined in Appendix A.)
\begin{figure}
\centerline{\epsfysize=4cm\epsfbox{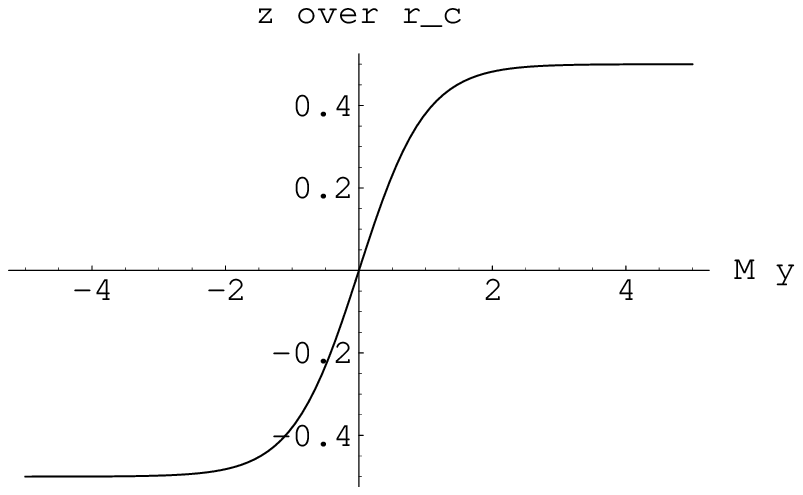}}
   \begin{center}
Fig.3\ Relation between two coordinates,
$z$ (compact) and $y$ (noncompact), (\ref{R.2}).
Vertical axis:\  $\frac{z}{r_c}$; 
Horizontal axis: $My$. 
   \end{center}
\end{figure}
In the "wall region" $|y|\ll 1/M$ (or $|z|\ll r_c$), 
$z$-coordinate and 
$y$-coordinate are almost same except a simple factor\ :\ 
$z\approx y\times\frac{Mr_c}{2}$. 
In the asymptotic (infrared) regions $|y|\gg 1/M$ (or $|z|\approx r_c/2$), 
they differ significantly\ :\ 
$z/r_c\approx \pm(\half-\e^{-2M|y|})$ as $My\ra \pm\infty$. 
Without confusion, we may take
\begin{eqnarray}
{r_c}=1\pr
\label{R.3}
\end{eqnarray}
(When $r_c$-dependence is required, it is easily obtained
by the substitution $z\ra z/r_c$.
)
In terms of the new coordinate $z$, the line element (\ref{R.1})
reduces to
\begin{eqnarray}
{ds}^2=\e^{-2\si(z)}\eta_\mn dx^\m dx^\n+
\frac{4}{(1-4z^2)^2}{dz}^2\com\nn
dz=\half (1-4z^2)dy\com\q
z\in (-\half,\half)\pr
\label{R.4}
\end{eqnarray}
$z=\pm \half$ are the points of the {\it coordinate singularity}
. The 5D Riemann scalar curvature is given by
\begin{eqnarray}
\Rhat=F (-2F{\si}''+5F{\si'}^2-2F'\si')\com\q
F\equiv 1-4z^2\com\q \si'=\frac{d\si}{dz}\com
\label{R.4B}
\end{eqnarray}
and it turns out that there are {\it no curvature singularities
anywhere}, for the solution we will consider. 
Let us consider the case the 5D Higgs field $\Phi(z)$ has the
following boundary condition(Fig.4).
\begin{figure}
\centerline{\epsfysize=4cm\epsfbox{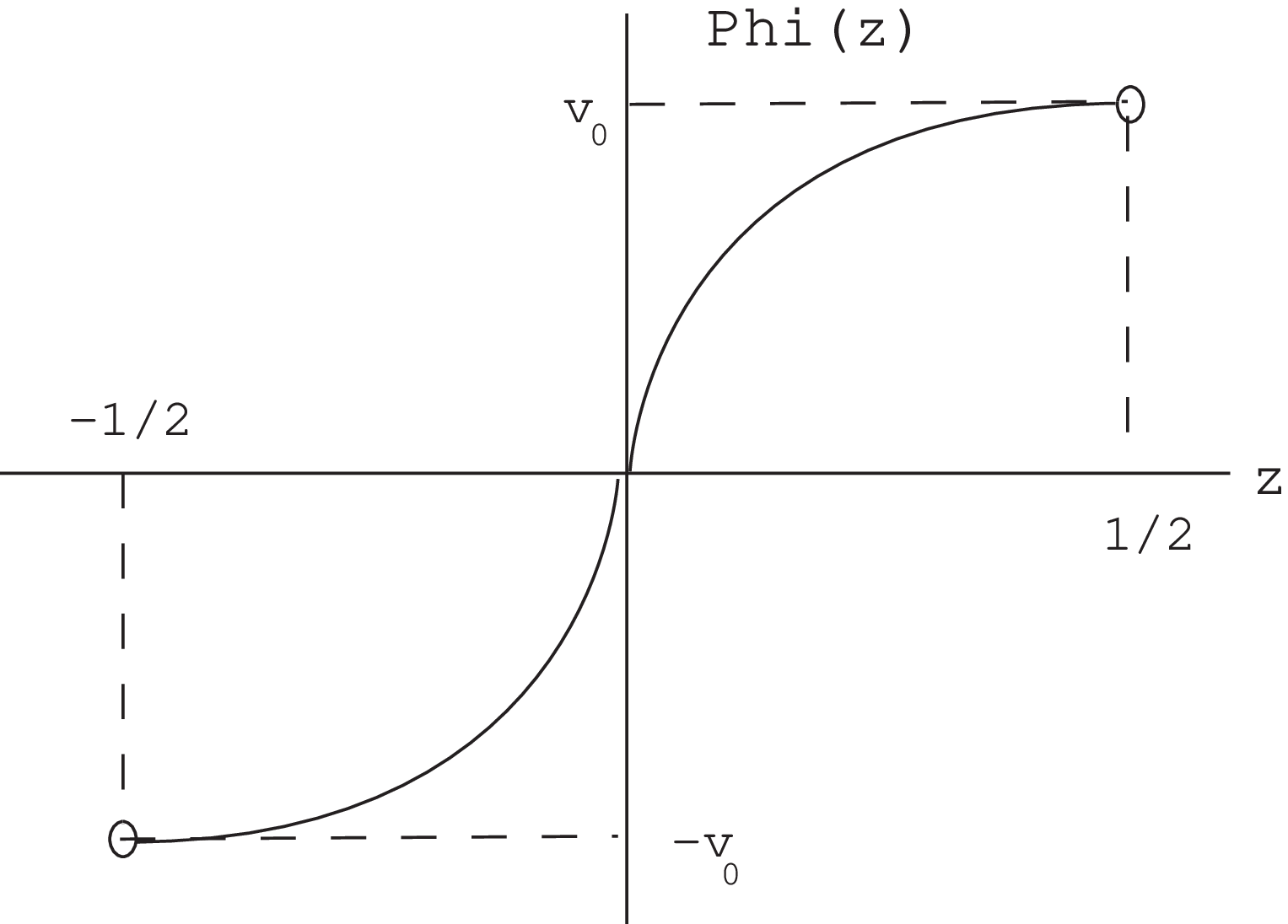}}
   \begin{center}
Fig.4\ The Higgs field with the boundary condition
(\ref{R.5}). Vertical axis:\  $\Phi(z)$; 
Horizontal axis: $z$. 
   \end{center}
\end{figure}
\begin{eqnarray}
\lim_{z\ra\pm(\half -0)}\Phi(z)\ra\pm v_0\com\q
v_0>0\pr
\label{R.5}
\end{eqnarray}
$\pm v_0$ is the vacuum expectation value in the asymptotic
region $z\ra\pm (\half-0)$. 
%
%
\section{$S^1$ Extra Axis}
In this section
we move to the case where the extra axis is $S^1$
through the following procedure.
If we can properly regularize the coordinate
singularity at $z=\pm\half$ in (\ref{R.4}) 
(see the next section), the coordinate region
$-\half\leq z\leq \half$ can be 
extended to 
${\bf R}=(-\infty,\infty)$ as follows.
\begin{enumerate}
\item
We require the periodic boundary condition\ :\ 
\begin{eqnarray}
\Phi(z)=\Phi(z+1)\com\q
\si(z)=\si(z+1)\q (\mbox{or}\q\si'(z)=\si'(z+1))
\ ;
\label{S.1}
\end{eqnarray}
\item
Values of $\Phi$ at $z=\half +{\bf Z}$\ are {\it defined} as, 
\begin{eqnarray}
\Phi(\half+{\bf Z})\equiv 0
\ ;
\label{S.2}
\end{eqnarray}
\item
The universal covering space is taken to be the real number
space\ :\ 
\begin{eqnarray}
[-\half,\half]\times {\bf Z}=(-\infty,\infty)={\bf R}
\ ;
\label{S.3}
\end{eqnarray}
\end{enumerate}
where $\ {\bf Z}=\{ 0,\pm 1,\pm 2,\cdots\}$. See Fig.5
for the schematic behavior of $\Phi(z)$.
\begin{figure}
\centerline{\epsfysize=4cm\epsfbox{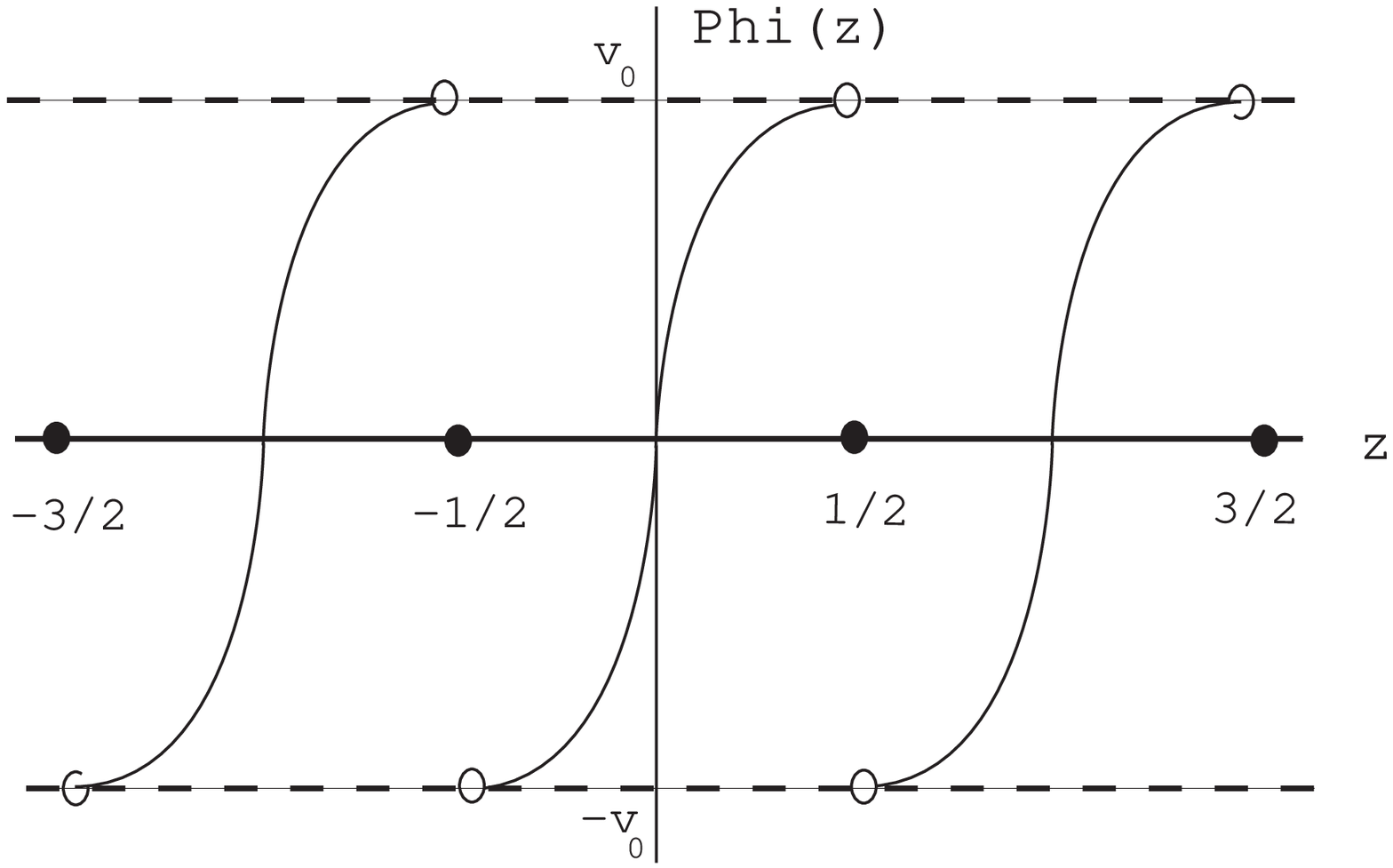}}
   \begin{center}
Fig.5\ The periodically extended Higgs field $\Phi(z)$. 
Vertical axis:\  $\Phi(z)$; 
Horizontal axis: $z$. 
   \end{center}
\end{figure}
Note that we have here {\it newly defined} the values of
$\Phi(z)$ at the singular points $z=\half+{\bf Z}$
(coordinate singularity, not the curvature singularity).
The points correspond to $y=\pm\infty$ of the original
coordinate, $y$. $\Phi(y=\pm\infty)$ are not defined
in Sec.3. In (\ref{S.2}) we have specified the
present treatment of $y=\pm\infty$, and which should be
regarded as a part of the present
infrared regularization.
\footnote{
This procedure reminds us of the similar one in the case of
making the sphere topology (compact) from the ${\bf E}^2$
space (non-compact) by introducing the point of infinity.
}
This specification turns
out to be important in Sec.5.

We furthermore note that the translation invariance $y\ra y+c$
in (\ref{R.1}) reduces to the periodicity invariance
(discrete version of the translation) $z\ra z+1$.
(The situation is the same as the lattice regularization
of the continuum space.) The lost of the translation freedom
$c$ is traded with the freedom of the coordinate choice $r_c$.

$\Phi(z)$ defined above has the properties:
\begin{description}
\item[P1]
Piecewise continuous. ({\it Discontinuous} at $z=\half+{\bf Z}$.) 
\item[P2]
Piecewise smooth. ($\Phi'(z)$ is piecewise continuous.)
\item[P3]
Periodic ($S^1$-symmetry )\ :\ $z\ra z+1$.
\item[P4]
Odd function of $z$ ($Z_2$-symmetry)\ :\ $\Phi(z)=-\Phi(-z)$.
\end{description}
We call these "$\Phi$-properties".

At present, except for the coordinate-singularity points
$z=\half+{\bf Z}$,
the metric (\ref{R.4}) is defined for 
$z\in {\bf R}=(-\infty,\infty)$:
\begin{eqnarray}
(G_{MN})=\left(
\begin{array}{cc}
\e^{-2\si(z)}\eta_\mn & 0 \\
0 & \frac{4}{{F(z)}^2}
\end{array}
         \right)
\com\label{S.4}
\end{eqnarray}
where $F(z)=1-4z^2$ at this stage (soon redefined)
and $M=(\m,z)$. 
The Einstein equation (\ref{int2}) reduces to
the following two coupled differential equations
for $\Phi(z)$ and $\si(z)$.
\begin{eqnarray}
-\frac{3}{2}{\si'}^2{F(z)}^2=-\frac{1}{8}{F(z)}^2{\Phi'}^2
+V\com\nn
\frac{3}{4}\frac{1}{F(z)}(\si'F(z))'
=\fourth{\Phi'}^2
\com\label{S.5}
\end{eqnarray}
where 
$\si'=\frac{d\si}{dz},\ {\Phi}'=\frac{d\Phi}{dz}$. 
In order to make the above equations periodic in
z\ :\ $z\ra z+1$, we must replace $F(z)=1-4z^2$ by its
"periodic generalization"
\footnote{
When we take another compact coordinate $w$,
defined in (\ref{sc1}), this "periodic generalization" is
not necessary.
}
:
\begin{eqnarray}
F(z)& = & \left\{
\begin{array}{c}
1-4z^2\q\mbox{for}\q -\half\leq z\leq\half\\
\\
1-4{z'}^2\q\mbox{for}\q  |z|>\half\\
(z=z'+n\ ,\ -\half\leq z'\leq\half\ ,\ n=\pm 1,\pm 2,\cdots)
\end{array}
     \right.
     \nn
&\equiv & [1-4z^2]\com
\label{S.6}
\end{eqnarray}
See . We denote the periodically generalization of a
function $f(x)$ as $[f(x)]$.
\begin{figure}
\centerline{\epsfysize=4cm\epsfbox{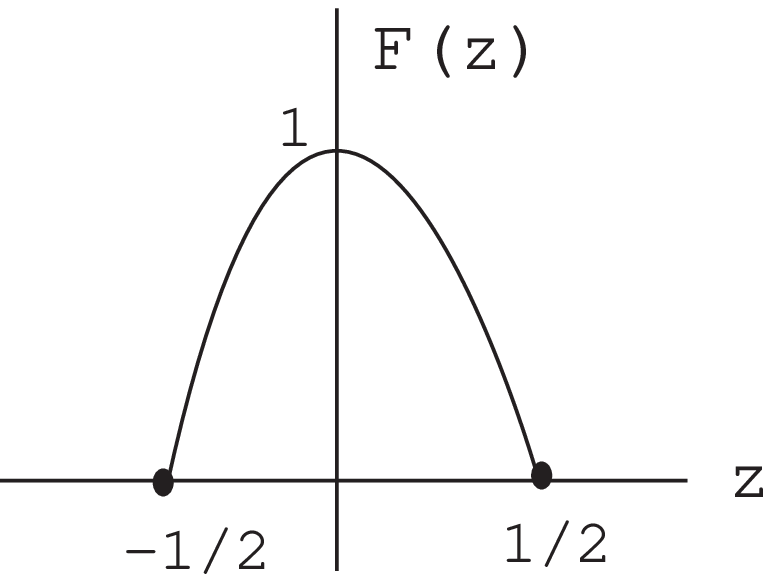}}
\centerline{\epsfysize=4cm\epsfbox{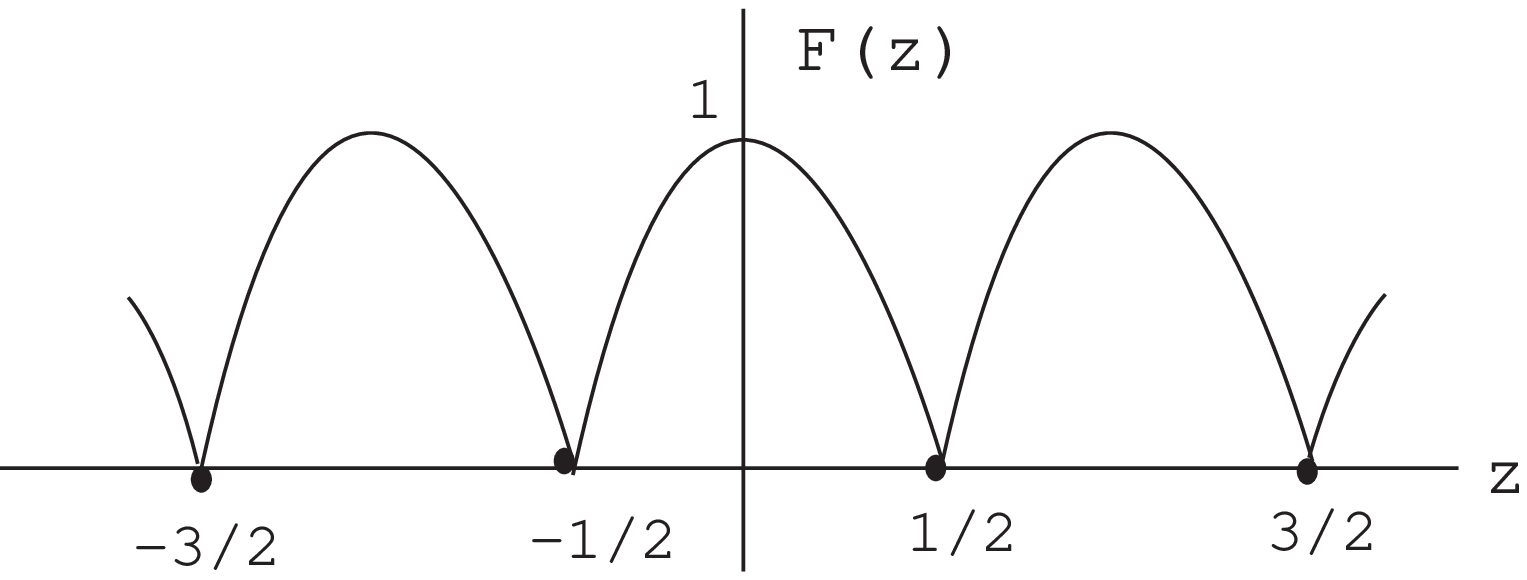}}
   \begin{center}
Fig.6\ [Above] The function $1-4z^2 (-1/2\leq z\leq 1/2)$;\ 
[Below] The periodically generalized function $[1-4z^2]$,
(\ref{S.6}). 
Horizontal axis: $z$. 
   \end{center}
\end{figure}
From the definition, $F(z)=[1-4z^2]$ has the following properties.
\begin{description}
\item[F1]
Continuous function of $z$.
\item[F2]
Piecewise smooth ($F'(z)$ is singular at $z=\half +{\bf Z}$).
\item[F3]
Periodic ($S^1$-symmetry) \ :\ $F(z)=F(z+1)$.
\item[F4]
Even function of $z$ ($Z_2$-symmetry)\ :\ $F(z)=F(-z)$.
\item[F5]Positive semi-definiteness\ :\ 
$F(z)\geq 0$.
\end{description}
We call these properties "F-properties".
This process of replacing $(1-4z^2)$ by $[1-4z^2]$ of (\ref{S.6})
should be regarded as a part of the present (infrared) 
regularization.

For some later use, we present here the {\it Fourier expansion}
of $F(z)$.
\begin{eqnarray}
F(z)=\frac{2}{3}+\frac{4}{\pi^2}\sum_{\l=1}^{\infty}
\frac{(-1)^{\l+1}}{\l^2}\cos(2\pi \l z)
\pr\label{S.7}
\end{eqnarray}
This expression can be taken as the definition of
$F(z)$ instead of (\ref{S.6}).

We can fix the asymptotic form of $\si'(z)$ using (\ref{S.5})
as follows.
In this process of "periodic generalization", the boundary condition
(\ref{R.5}) is also generalized to be
\begin{eqnarray}
\lim_{z\ra\pm(\half -0)+\bfZ}\Phi(z)=\pm v_0\com\q 
v_0>0\pr
\label{S.8}
\end{eqnarray}
(Note $\Phi(\pm\half+\bfZ)=0$ as introduced in (\ref{S.2}).)
We call the asymptotic regions $\{z|z\ra\pm(\half -0)+\bfZ\}$
IR-regions and another regions $\{z|z\ra\pm 0+\bfZ\}$
UV-regions. 
In the IR-regions, 
$\Phi'\ra 0$ from the above equation, therefore
$\si'F(z)\ra$ constant from the second eq. of (\ref{S.5}).
Furthermore, using the first equation, we obtain
\begin{eqnarray}
\lim_{z\ra\pm(\half -0)+\bfZ}\Si(z)=\pm \om\com\q
\om=\sqrt{-\frac{2\La}{3}}\pr
\label{S.9}
\end{eqnarray}
where $\Si(z)\equiv F(z)\si'(z)$. 
Note that $\Si(z)=2\frac{d\si}{dy}$ for $-\half\leq z\leq\half$.
The behavior of the "warp" factor field $\Si(z)$ will
be shown to be similar to the Higgs field $\Phi(z)$. 
($\Phi(z)$ and $\Si(z)$ will be parallelly discussed in
Sec.5 and 6.) 
In eq.(\ref{S.9}), 
we notice $\La$ should be negative\ :\ $\La\leq 0$.
From the first eq. of (\ref{S.5}), we know
\begin{eqnarray}
\frac{1}{8}{F(z)}^2{\Phi'}^2=
\frac{3}{2}{\Si(z)}^2+V\geq 0
\pr\label{S.10}
\end{eqnarray}
From the field equations (\ref{S.5}) and the boundary conditions
(\ref{S.8},\ref{S.9}), we conclude $\Si(z)$ and $\Phi(z)$
are {\it odd} functions of $z$. Hence we have
\begin{eqnarray}
\Si(z)=0 \q\mbox{and}\q \Phi(z)=0\ \mbox{at}\ 
z=0
\pr\label{S.11}
\end{eqnarray}
(This condition will be used to solve the 
field equations (\ref{S.5}) 
numerically. The boundary conditions (\ref{S.8},\ref{S.9})
cannot be taken due to the singularity. See Appendix B.) 
Applying this result to (\ref{S.10}), we know
$\frac{\la}{4}{v_0}^4+\La\geq 0$. Combining the previous
result, we obtain\cite{SI00apr}
\begin{eqnarray}
-\frac{\la}{4}{v_0}^4\leq\La\leq 0
\pr\label{S.12}
\end{eqnarray}
It says the sign of $\La$ should be negative (anti de Sitter)
and the absolute value has the {\it upper} bound:\ 
$|\La|\leq\frac{\la}{4}{v_0}^4$ .
Near the singular points, $z\ra \pm(\half-0)$,
the asymptotic behavior of the line element is given by,
from (\ref{S.9}),
\begin{eqnarray}
ds^2=\left|\frac{1+2z}{1-2z}\right|^{\mp\om/2}\eta_\mn dx^\m dx^\n
+\frac{4}{(1-4z^2)^2}dz^2\com\q
\om=\sqrt{-\frac{2\La}{3}}\com
\label{S.13}
\end{eqnarray}
which shows the singular points are 
horizons
.
The power-law of the Weyl factor ("warp factor")
indicates the {\it scaling} behavior of the system
when $z\ra \pm(\half-0)$. See Sec.7.

%
%
\section{Infrared Regularization at $z=\half+\bfZ$}
We come to the most important part of the present regularization.
Before the presentation, we give here a mathematically well-known
fact. The periodic step function $\th(x)$ defined by
\begin{eqnarray}
\th(x)=\left\{
\begin{array}{cc}
1\q & 2n\ep<x<(2n+1)\ep \\
0\q & x=n\ep \\
-1\q & (2n+1)\ep<x<(2n+2)\ep
\end{array}
       \right.
\com\label{reg1}
\end{eqnarray}
where $n\in\bfZ$ (see Fig.7).
It has the following properties:\ 
\begin{description}
\item[T1]
Piecewise continuous. ({\it Discontinuous} at $x\in \ep\times\bfZ $.)
\item[T2]
Piecewise smooth.
\item[T3]
Periodic ($S^1$-symmetry)\ :\ $\th(x+2\ep)=\th(x)$.
\item[T4]
Odd function of $x$ ($Z_2$-symmetry)\ :\ $\th(x)=-\th(-x)$.
\item[T5]
Symmetric with respect to the axes $x=(\pm\half+2\bfZ)\ep$
\ (UV-IR symmetry). 
\end{description}
These are similar to $\Phi$-properties (for $\ep=\half$) except 
for Property T1\ (the number of discontinuous points doubles) 
and Property T5\ (UV-IR relation).
We call these properties "$\th$-properties".
The periodic step function $\th(x)$, which is {\it discontinuous}, 
has the following Fourier expansion.
\begin{eqnarray}
\th(x)=\frac{4}{\pi}\sum_{\l=0}^\infty\frac{1}{2\l+1}
\sin\{(2\l+1)\pi\frac{x}{\ep}\}
\pr\label{reg2}
\end{eqnarray}
(Compare the Fourier expansion of the {\it continuous}
function $F(z)$, (\ref{S.7}).
Main changes are, $2\l$ in $F(z)$ is replaced by 
$(2\l+1)/\ep=2(2\l+1)$
for $\ep=1/2$, and the coefficient $(-1)^{\l+1}/\l^2$ is by $1/(2\l+1)$.
The discontinuous case is less convergent series than the continuous
case.
)
When we {\it regularize} (\ref{reg2}) by the {\it finite} ($L$) sum,
\begin{eqnarray}
\th_L(x)=\frac{4}{\pi}\sum_{\l=0}^L\frac{1}{2\l+1}
\sin\{(2\l+1)\pi\frac{x}{\ep}\}
\com\label{reg3}
\end{eqnarray}
then 
$\th_L(x)$ has the following new properties compared with $\th(x)$:\ 
\begin{description}
\item[TL1]
{\it Continuous} everywhere $x\in \bfR=(-\infty,\infty)$
(see Fig.5). Especially 
$\th_L(x\in \ep{\bf Z})=0$.
\item[TL2]
Smooth everywhere.
\end{description}
Other items 3,4 and 5 are the same as $\th(x)$:\ 
{\bf TL3}={\bf T3}, {\bf TL4}={\bf T4}, {\bf TL5}={\bf T5}.
We call these properties "$\th_L$-properties". 
This simple example characteristically shows
that a {\it discontinuous} function can be naturally regularized
by a {\it continuous} function by {\it truncating} the {\it infinite}
Fourier series by a {\it finite} $L$ sum. $L$ is here regarded
as an infrared {\it regularization parameter}.
The continuousness is indispensable for 
a wall-configuration with finite thickness or for
a well-defined regularization.
The meaning of $1/L$ is the "thickness" of the walls or
anti-walls of
${\th_L}'(x)$ at $x=\ep\bfZ$.
\footnote{
In the "wall region" around the origin 
$|\frac{x}{\ep}|\ll 1$, all L+1 terms equally
dominate in the RHS of (\ref{reg3})\ :\ 
$\th_L(x)\approx\frac{4}{\pi}\sum_{\l=0}^L\frac{1}{2\l+1}
\{(2\l+1)\pi\frac{x}{\ep}\}=\frac{4}{\ep}(L+1)x$.
Therefore the thickness $w$ can be defined as\ :\
$\frac{4}{\ep}(L+1)\frac{w}{2}=\half$,hence we have
$w=\frac{\ep}{4(L+1)} $.
The same thing can be said about all "wall-regions" 
around $x\in 2\ep \bfZ$ and about all 
"anti-wall regions" around $x\in \ep (2\bfZ+1)$.
It is well-known that, in these "wall and anti-wall regions"
the truncated function $\th_L(x)$ most deviate from $\th(x)$
because the neglected terms(high-frequency modes) begin to
equally contribute with low-frequency ones (Gibbs's phenomenon).
}
The thickness here is purely a {\it regularization} effect.
See Fig.7.
\begin{figure}
\centerline{\epsfysize=4cm\epsfbox{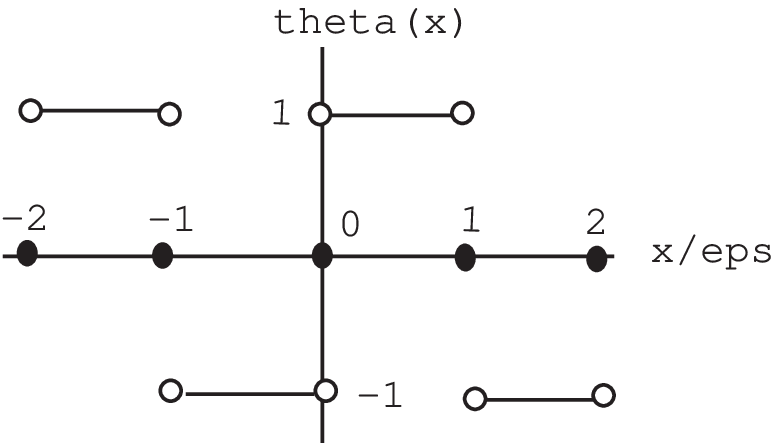}}
\centerline{\epsfysize=4cm\epsfbox{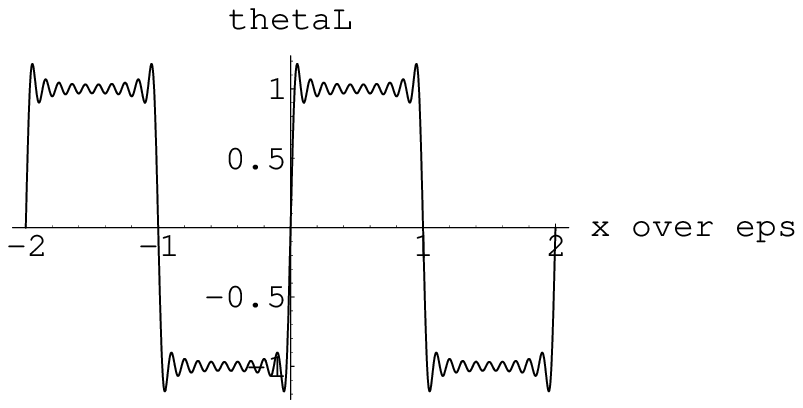}}
   \begin{center}
Fig.7\ 
[Above]\ The periodic step function $\th(x)$ ((\ref{reg1}) or (\ref{reg2}))
and [Below]\ its regularized function $\th_{L=9}(x)$ ((\ref{reg3})). 
Horizontal axes: $x/\ep$. 
   \end{center}
\end{figure}

With the above fact in mind, we propose here 
a {\it new regularization} in order to treat the singularity
at $z=\half+\bfZ$ of the solutions in the previous section. 
First we know $\Phi(z)$ and
$\Si(z)=F(z)\si'$ behave like a periodic $\th$-function
at some parameters limit ( the infinitely-thin wall
limit). 
\footnote{
The infinitely thin wall limit is given by $\la {v_0}^2\ra +\infty$
(for a given $v_0$ and an appropriately chosen $\La$).
}
Both satisfy the $\Phi$-properties in
Sec.4. Imitating (\ref{reg3}) with $\ep=\half$, 
we take, as the  regularized 
solution
of (\ref{S.5}), the following forms for
$\Phi(z)$ and $\Si(z)$. 
\begin{eqnarray}
\Phi_L(z)=v_0\frac{4}{\pi}\sum_{\l=0}^L\frac{d_\l}{2\l+1}
\sin\{(2\l+1)2\pi z\}
\com\nn
\Si_L(z)=\om\frac{4}{\pi}\sum_{\l=0}^L\frac{c_\l}{2\l+1}
\sin\{(2\l+1)2\pi z\}
\com
\label{reg4}
\end{eqnarray}
where $c's$ and $d's$ are some numbers to be determined
appropriately and $L$ is
the new {\it regularization parameter} which should be
taken sufficiently large.
At present, the way to fix the coefficients,
$c's$ and $d's$, relies on
a numerical method. (See App.B.1 for solving (\ref{S.5})
numerically, and see App.B.2 for fixing the coefficients
by the least square method.) 
As explained in App.C, 
the behavior of $\{d_\l\}$ and 
$\{c_\l\}$ has three "phases":
\ i) $2\l+1\ll 1/(4\pi w_{UV})$,
ii) $2\l+1\approx 1/(4\pi w_{UV})$, and
iii) $2\l+1\gg 1/(4\pi w_{UV})$,
where $w_{UV}$ is the thickness around the UV regions ($z\sim \bfZ$).
The critical value $\l=L^* $ is
given by the vacuum parameters:
\begin{eqnarray}
\frac{L^*}{r_c}\sim\frac{1}{8\pi w_{UV}}
\sim\sqrt{\la\vz^2}=m_H 
\com
\label{reg4.5}
\end{eqnarray}
where $\sqrt{\la\vz^2}$ is identified as the (5 dim)
Higgs mass $m_H$ defined by
${m_H}^2\equiv \half V''(\vz)$. 
The (length) scale $w_{UV}$ is an important quantity
in the mass hierarchy problem.
\footnote{
This new mass scale
$L^*/r_c$ corresponds to the parameter $k$ in the original
RS model\cite{RS9905,RS9906}.
}
These values $L^*$ and $w_{UV}$ are {\it independent of the regularization
parameter} $L$.  
This point should be compared with the thickness appeared in $\th_L(x)$.
The condition for the {\it dimensional reduction}, from 5 dim
to 4 dim, is given by
\begin{eqnarray}
L^*\sim\frac{r_c}{8\pi w_{UV}}\gg 1
\pr
\label{reg4.55}
\end{eqnarray}
In the present regularized solution (\ref{reg4}), 
the {\it UV-IR symmetry}
(i.e., symmetry w.r.t the axes $z=\pm\fourth +\bfZ$)
holds. Therefore another width $w_{IR}$ around the
IR-regions ($z\sim \half+\bfZ$) is the same as $w_{UV}$.
\begin{eqnarray}
w_{IR}=w_{UV}
\pr
\label{reg4.6}
\end{eqnarray}
$w_{IR}$ is protected against $L$(regulator) dependence, at least,
taking the regularized form of (\ref{reg4}).
(This should be compared with the $\th_L(x)$ case, where 
$w_{IR}=w_{UV}\sim L^{-1}$.)

Both $\Phi_L$ and $\Si_L$ have 
the $\th_L$-properties with $\ep=\half$.
We consider the case that $L^*$ is 
large($L\gg L^*\gg 1$), that is, the solutions $\Phi(z)$ and $\Si(z)$ are near the $\th$-function. The infinitely-thin
wall limit ($\th$ or $\del$-function distribution)
corresponds to, in (\ref{reg4}), the following case.
\begin{eqnarray}
L \gg L^* \ra \infty\com\q
d_\l\ra 1\com\q c_\l\ra 1\q\mbox{for all $\l$'s}
\pr
\label{reg5}
\end{eqnarray}
Taken into account the condition that 
the 5D {\it classical} Einstein equation works, that is, 
the new mass scale
$L^*/r_c$ should be much less than 5D Planck scale $M$, 
we should have
the following relations between parameters.
\footnote{
In (\ref{reg6}) the $r_c$-dependence is explicitly
written. The eq.(\ref{R.3}), which was introduced purely for the
notational simplicity, should be taken off and, instead, 
$Mr_c\gg 1$ should be considered.
}
\begin{eqnarray}
\frac{1}{r_c}\ll \frac{L^*}{r_c}\ll M
\pr
\label{reg6}
\end{eqnarray}
We regard the above three parameters,
$M$(fundamental scale),$r_c$(compactification size) and
$L^*$(wall-thickness parameter), as the {\it fundamental parameters}
of the theory.

The numerical results of (\ref{reg4}) 
are given in the next section.
(See App.B for further detail of the calculation.)
The solution (\ref{reg4}) is the regularized solution of (\ref{S.5}), 
not a true one . 
(How to improve (\ref{reg4}) perturbatively, 
in order to approach a true solution,  
is proposed in Sec.8. 
In the practical and numerical point of view, the solution
(\ref{reg4}) is sufficiently close to the true
solution.) It is, however, sufficient to claim 
the existence of the solution of (\ref{S.5})
that has the wall-anti-wall (kink-anti-kink)
configuration.

%
%
\section{Final Numerical Result of $S^1/Z_2$ Compactification}

In Fig.8,9 and 10, we plot three sample solutions of 
(\ref{reg4}) corresponding to the following three vacua respectively.
The configurations approach to the $\th$-function  in the
order of Vac.1, 2 and 3. 
\begin{eqnarray}
\mbox{Vacuum 1}\q \la=20.0\mbox{(input)}\ ,\ 
v_0=1.0\mbox{(input)}\ ,\ \La=\ -1.88855\ \nn
(\om=1.12207 )\ ,
L=19\mbox{(input)};\ 
w_{UV}\sim 8.9\times 10^{-3} (L^* \sim 4.5).
\label{res1a}\\
\mbox{Vacuum 2}\q \la=40.0\mbox{(input)}\ ,\ 
v_0=1.0\mbox{(input)}\ ,\ \La=\ -3.77762\ \nn
(\om=1.58695)\ ,
L=19\mbox{(input)};\ 
w_{UV}\sim 6.3\times 10^{-3} (L^* \sim 6.3).
\label{res1b}\\
\mbox{Vacuum 3}\q \la=100.0\mbox{(input)}\ ,\ 
v_0=1.0\mbox{(input)}\ ,\ \La=\ -9.4440537\ \nn
(\om=2.5091903 )\ ,
L=19\mbox{(input)};\ 
w_{UV}\sim 4.0\times 10^{-3} (L^* \sim 10.).
\label{res1c}
\end{eqnarray}

\begin{figure}
\centerline{\epsfysize=50mm\epsfbox{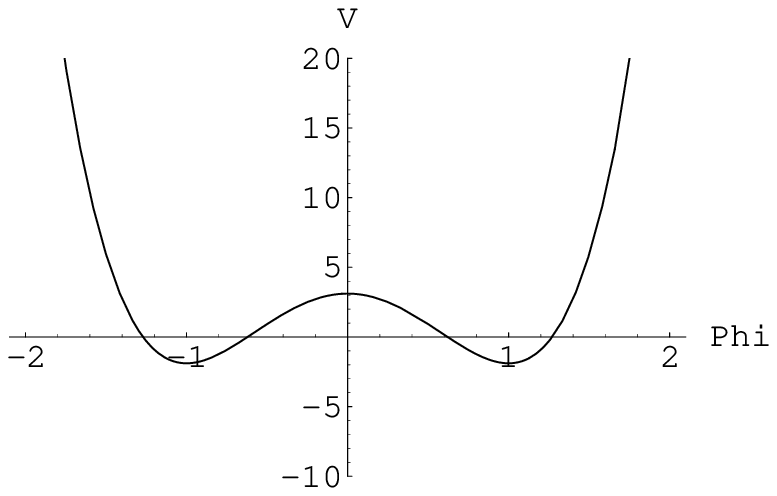}}
\centerline{\epsfysize=50mm\epsfbox{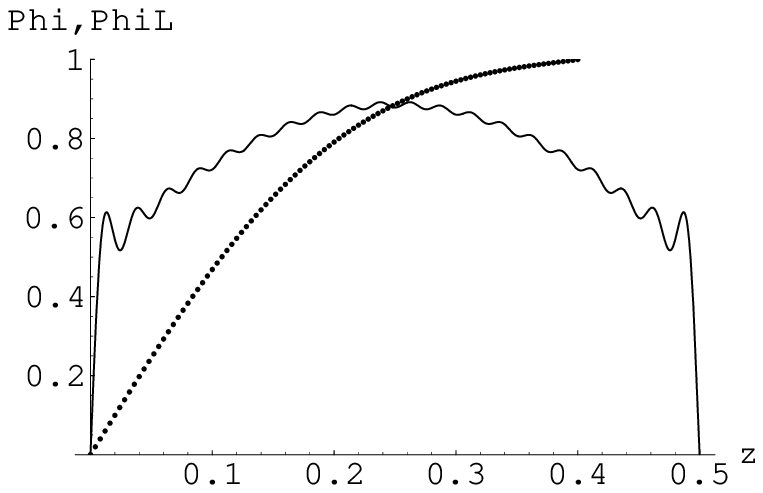}}
\centerline{\epsfysize=50mm\epsfbox{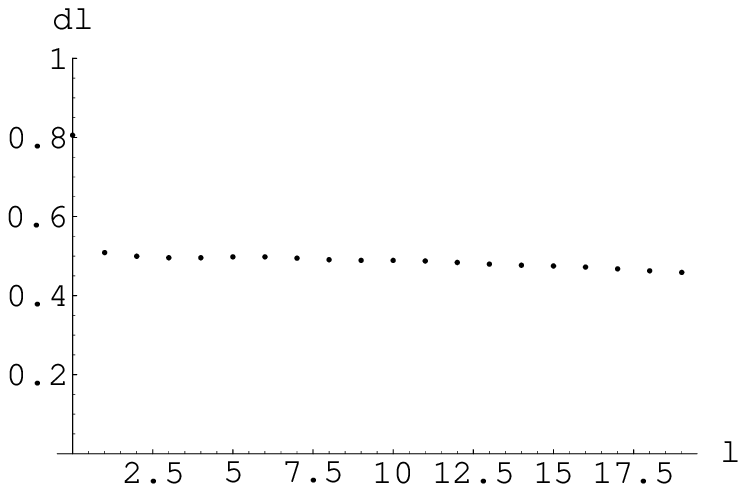}}
   \begin{center}
Fig.8\ Vacuum 1 $(\la,\vz,\La)=(20.0,1.0,-1.88855)$:
\ [Top] Higgs Potential $V(\Phi)$, (\ref{int1}). 
The horizontal axis is $\Phi$. 
\ [Middle] Numerical result of $\Phi(z)$ 
(dotted points, $0\leq z\leq 0.4$) for (\ref{S.5}) 
and its least-square fit solution $\Phi_L(z)$, 
(\ref{reg4}), ($0\leq z\leq 0.5$).
The horizontal axis is $z$. \newline
[Bottom] The coefficients \{$d_\l\ ;\l=0,1,\cdots,L=19$\} of $\Phi_L(z)$.
 The horizontal axis is $\l$. 
   \end{center}
\end{figure}

\begin{figure}
\centerline{\epsfysize=50mm\epsfbox{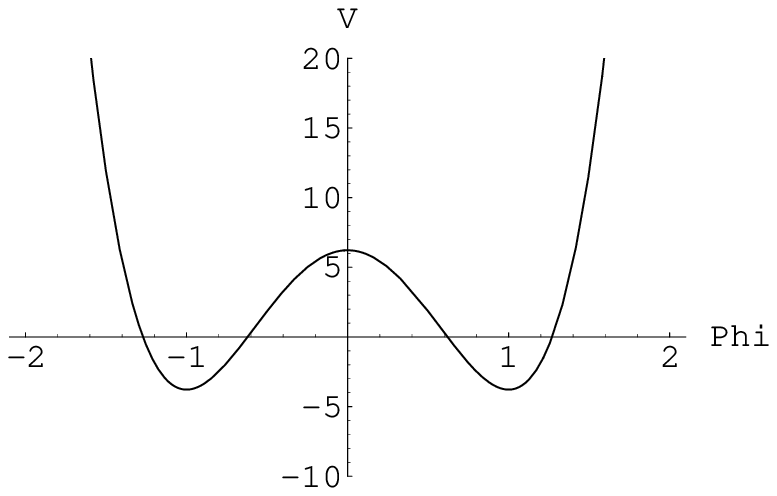}}
\centerline{\epsfysize=50mm\epsfbox{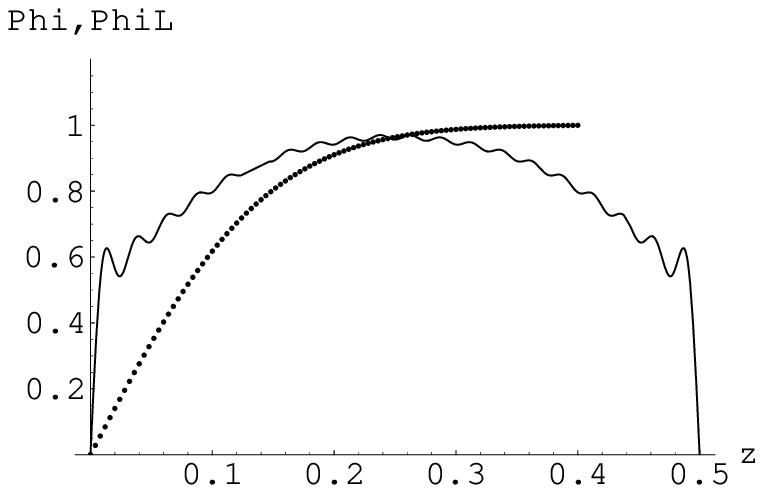}}
\centerline{\epsfysize=50mm\epsfbox{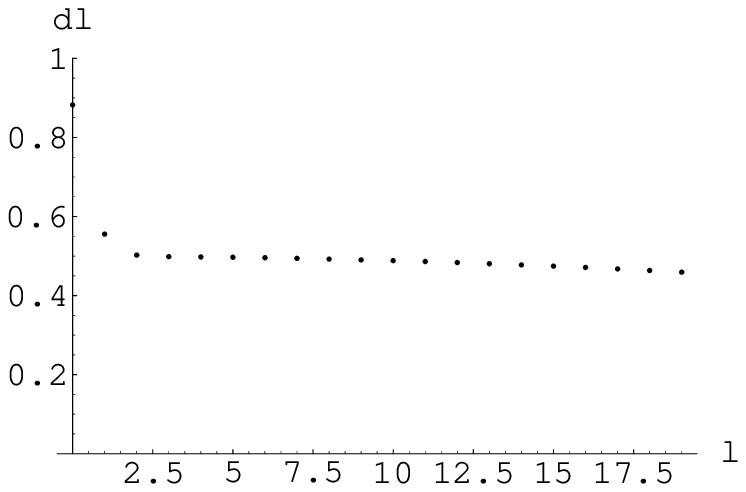}}
   \begin{center}
Fig.9\ Vacuum 2 $(\la,\vz,\La)=(40.0,1.0,-3.77762)$:
\ same as the figure caption of Fig.8.
   \end{center}
\end{figure}

\begin{figure}
\centerline{\epsfysize=50mm\epsfbox{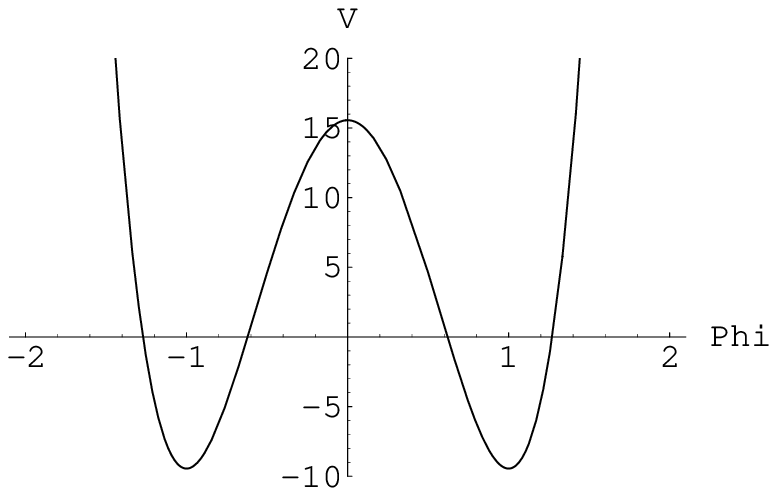}}
\centerline{\epsfysize=50mm\epsfbox{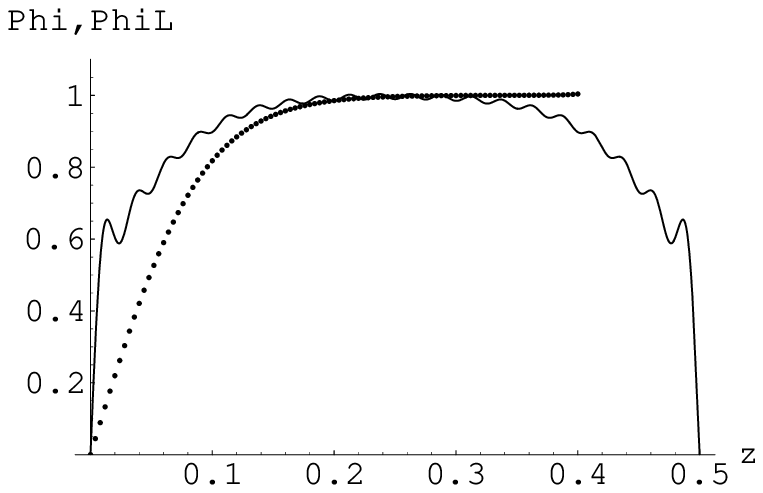}}
\centerline{\epsfysize=50mm\epsfbox{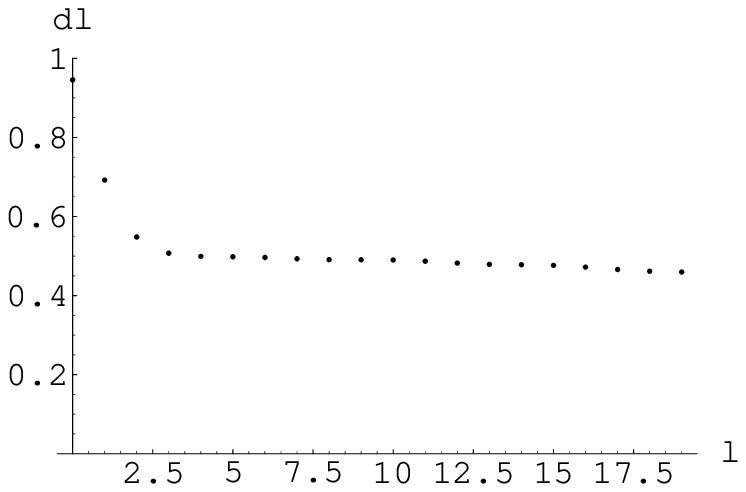}}
   \begin{center}
Fig.10\ Vacuum 3 $(\la,\vz,\La)=(100.0,1.0,-9.4440537)$:
\ same as the figure caption of Fig.8.
   \end{center}
\end{figure}

Note that the values of the (5D) cosmological term,
$\La$, are very finely chosen
so that the boundary conditions (\ref{S.8})
and (\ref{S.9}) are satisfied. 
All digits appearing are "significant figures".
As the
configuration approaches the $\th$-function limit
(${m_H}^2=\la\vz^2\ra \infty$), 
the necessary number of digits increases. 
(See also Vac.1w in App.A.) 
This shows the cosmological
constant is, for a given $\la$ and $\vz$, dynamically derived in the present
framework. Note that $\La$ is directly related with
the 4 dim cosmological constant\cite{SI00apr}.
\footnote{
The cosmological constant problem was reviewed in \cite{Weinb89}. 
One standard approach to the dynamical cosmological
constant is the radiative correction. 
Such scenario was explicitly done, using the Coleman-Weinberg
mechanism, in \cite{SI84}.
}
In App.A, we give another result for Vac.1 with
a different compactification coordinate.

The solutions of $\Si(z)$ and $\Si_L(z)$ are similar
to $\Phi(z)$ and $\Phi_L(z)$. 
In Fig.11, $\Si(z)$ and $\Si_L(z)$
are plotted for the case of Vacuum 3.
For each vacuum, $\Si(z)$ is always closer to the 
$\th$-function limit
than $\Phi(z)$. 

\begin{figure}
\centerline{\epsfysize=50mm\epsfbox{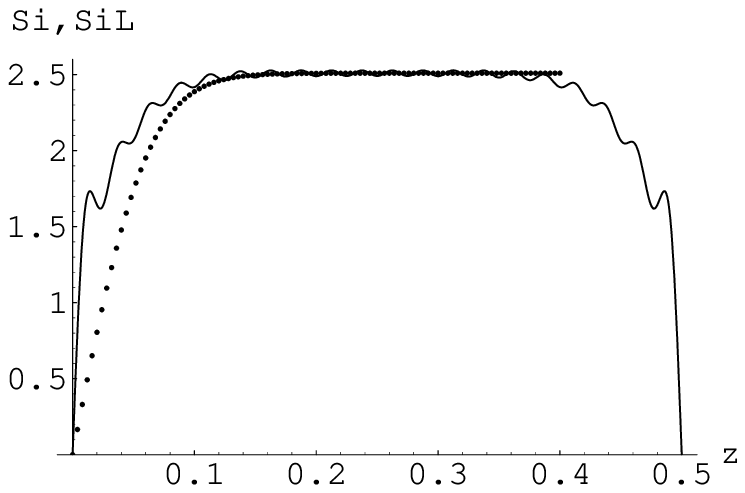}}
\centerline{\epsfysize=50mm\epsfbox{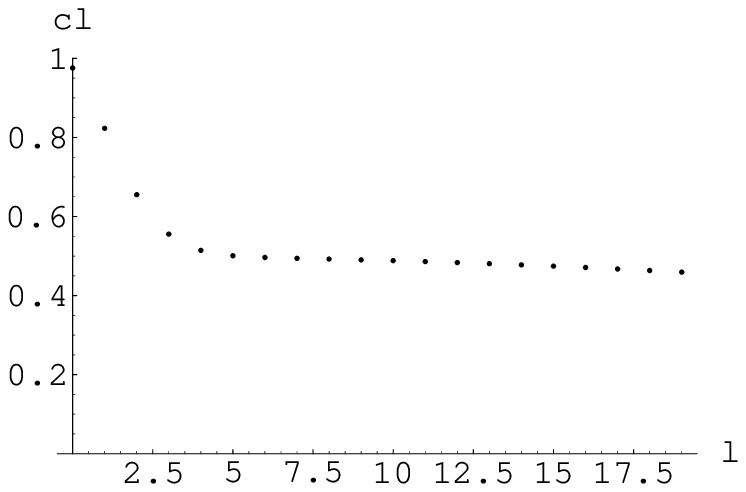}}
   \begin{center}
Fig.11\ Vacuum 3:
\ [Above] Numerical result of $\Si(z)$ 
(dotted points, $0\leq z\leq 0.4$)
and its least-square fit solution $\Si_L(z)$, 
(\ref{reg4}), ($0\leq z\leq 0.5$).
The horizontal axis is $z$. \newline
[Below] The coefficients \{$c_\l\ ;\l=0,1,\cdots,L=19$\} of $\Si_L(z)$.
 The horizontal axis is $\l$. 
   \end{center}
\end{figure}

The following items can be read from the above output data.
\begin{enumerate}
\item
As the Higgs potential has deeper valleies, 
which corresponds to the case that 
the 5D Higgs mass ($m_H=\sqrt{\la\vz^2}$)
becomes larger, $\Phi$ and $\Si$ approach 
the $\th$-function. All coefficients $d_\l$'s and $c_\l$'s
are expected to approach 1 (the limit of (\ref{reg5})).
\item
The wavy region, explained in (\ref{tra3}), is
not so clear in Vacuum 1-3, but can be seen in Vacuum 1w
of App.A.
\end{enumerate}

%
%
\section{Properties of the Solution}

\subsection{Wall-Anti-Wall Configuration}

In the previous section, the regularized (numerical)
solutions for the wall-anti-wall configuration
are given. 
$S^1/Z_2$ compactification is just taking the
segment $[0,\half]$ for the periodic coordinate $z$
with periodicity $1$.
The derivative of $\Phi_L(z)$ of Vacuum 3 (Fig.10)
is plotted in Fig.12. Here we see the present
approach surely gives the wall-anti-wall configuration.
The walls appear at {\it both ends} of the extra axis,
not at some middle points in the axis. The situation
is the same as that in the lattice domain wall\cite{Sha93,Vra98}.
We stress the points:\ 
1)\ the anti-wall is realized by the present
{\it IR regularization} where the discontinuity of $\Phi(z)$
and $\Si(z)$ at the singular points ($\half+\bfZ$)
is avoided by {\it truncating the infinite Fourier series};\ 
2)\ the stability of the solution is guaranteed by the boundary conditions;\ 
3)\ UV-IR symmetry is realized in the present form of the
regularized solution.
\footnote{
In ref.\cite{GS0005}, the wall-anti-wall configuration is
considered in a modified RS-model and UV-IR symmetry is
suggested.
}
We need not the radion field which was considered 
, for the stability, in \cite{RS9905}
and was developed in \cite{GW9907b}. (In the place of
the radion field $T^2(x)$(eq.(13) of \cite{RS9905}), 
from eq.(\ref{R.4}), the fixed function of $z$, 
$4/(1-4z^2)^2$, appears in the present scenario. )

\begin{figure}
\centerline{\epsfysize=50mm\epsfbox{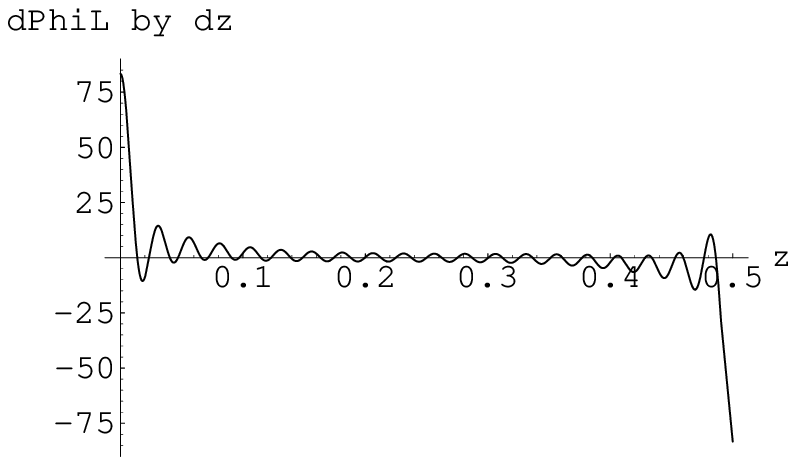}}
   \begin{center}
Fig.12\ 
Plot of $\frac{d\Phi_L(z)}{dz}$ where $\Phi_L(z)$ is
given in Fig.10(Vacuum 3).
The horizontal axis is $z$. 
   \end{center}
\end{figure}
\subsection{Brane Tensions at the Wall and the Anti-Wall}
We make a remark on a comparative aspect between the wall
and the anti-wall. Let us consider the "$\th$-function limit"
(thin wall limit) $m_Hr_c=\sqrt{\la\vz^2}r_c\sim L^*\gg 1$.
The behaviors of $\Phi(z)$ and $\Si(z)=\si'F(z)$ are 
considered to be the 
periodic step functions shown in Fig.13b and d.
Correspondingly those of $\Phi'$ and $\Si'$ can be written as
\begin{eqnarray}
\Phi'=\vz\sum_{n\in \bfZ}\{
\del(z-n)-\del(z-n-\half)\}\com\nn
\Si'=\om\sum_{n\in \bfZ}\{
\del(z-n)-\del(z-n-\half)\}
\pr
\label{concM1}
\end{eqnarray}
See Fig.13a and c. 
\begin{figure}
\centerline{\epsfysize=6.5cm\epsfbox{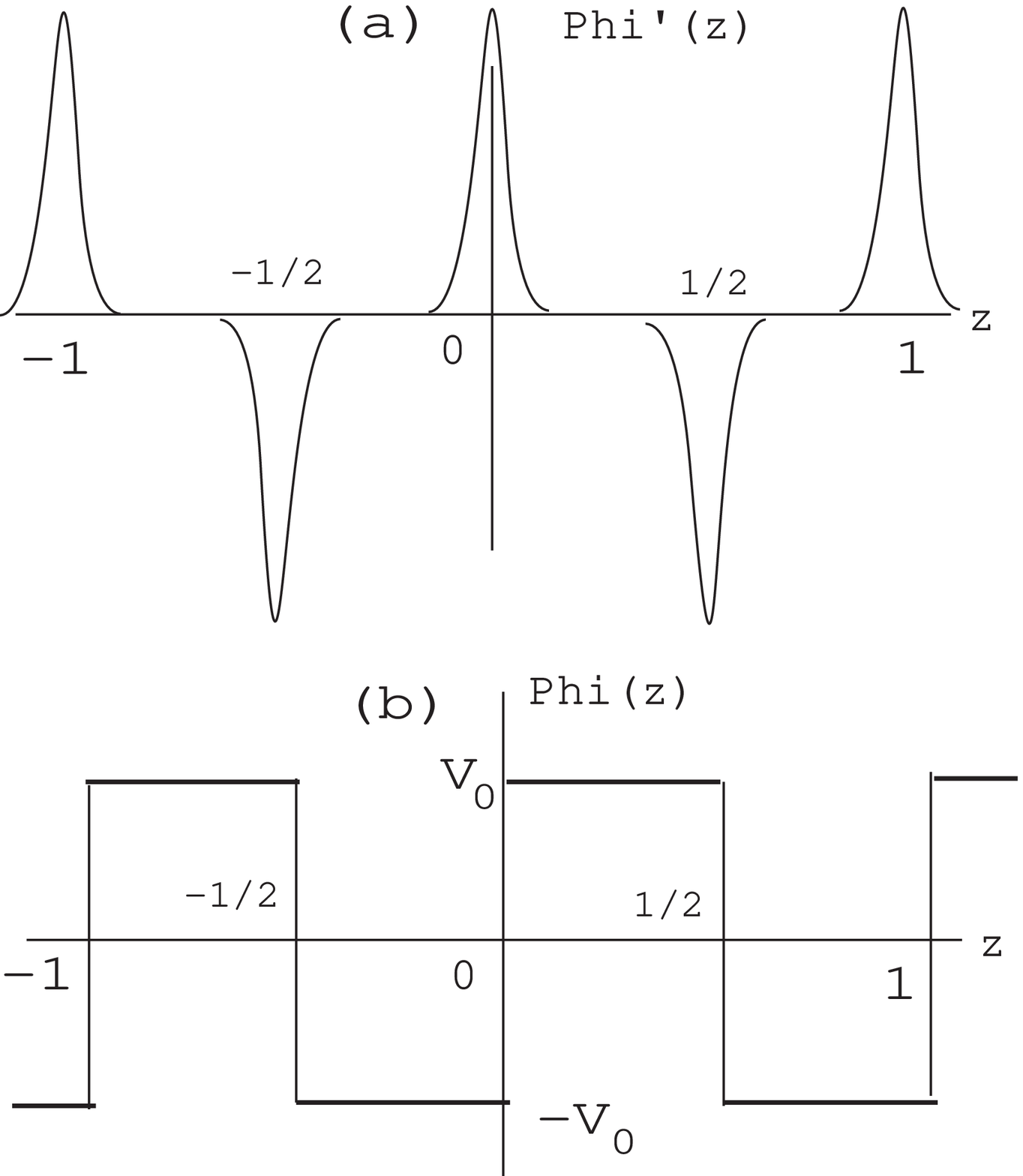}}
\centerline{\epsfysize=10cm\epsfbox{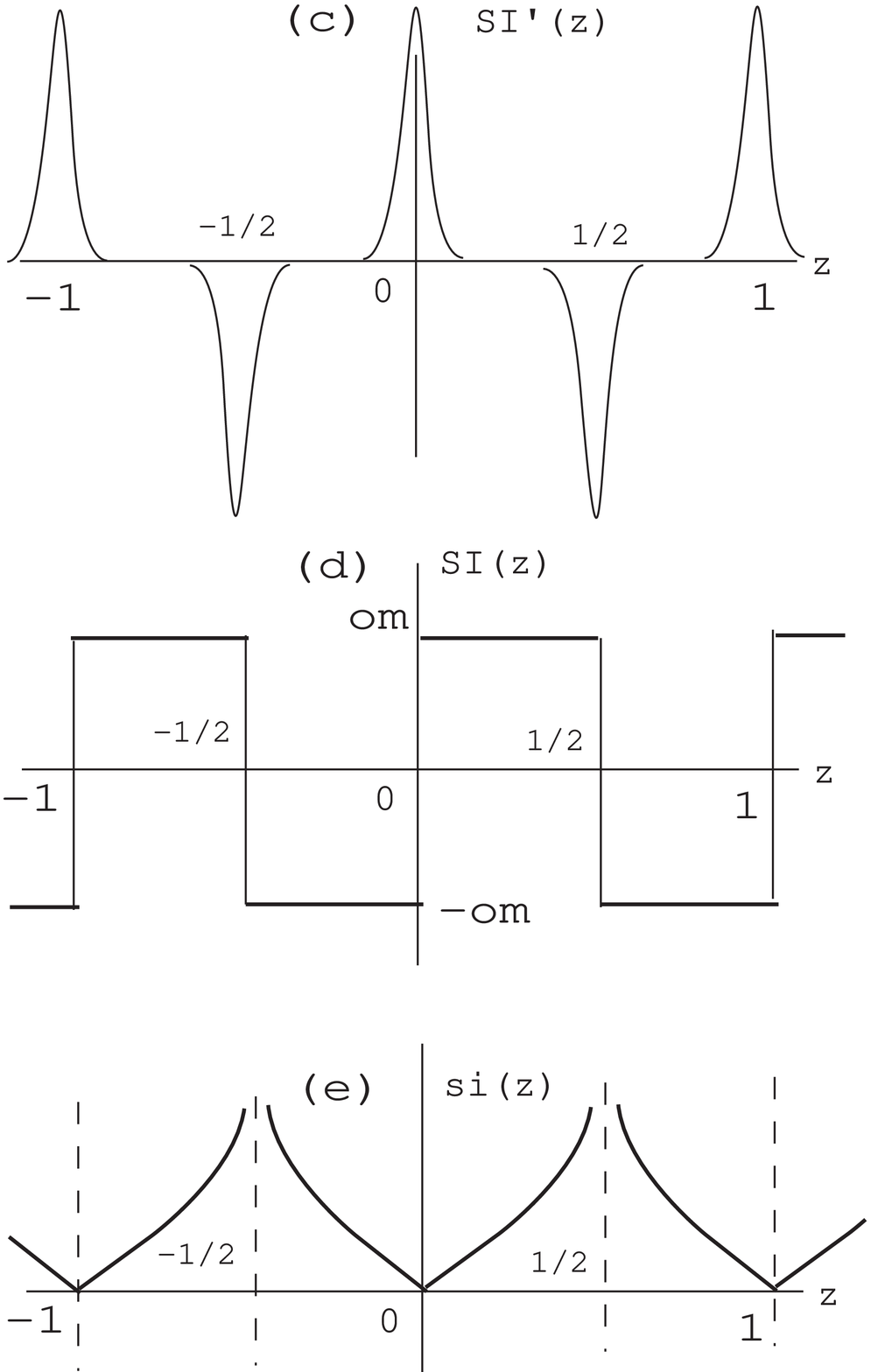}}
   \begin{center}
Fig.13\ Behaviors at the thin-wall limit. 
Horizontal axis: $z$. 
(a)\ $\Phi'(z)$;\ (b)\ $\Phi(z)$;\ 
(c)\ $\Si'(z)$;\ (d)\ $\Si(z)$;\ (e)\ $\si(z)$. 
Especially, in Fig.13(e), $\si\sim \om z$ for $z\sim +0$ and
$\si\sim (\om/4)\ln (2/(1-2z))+\mbox{const.}$
for $z\sim (1/2)-0$. 
   \end{center}
\end{figure}
The "warp" factor
$\si(z)$ behaves as in Fig.13e.
These behaviors should be compared with (8)-(10) of 
Ref.\cite{RS9905}. The parts of minus-delta-function in 
eq.(\ref{concM1})
correspond to the anti-walls. Now we evaluate the integrand
of the 5D action (\ref{int1}) near $z\approx 0$(Wall) and
$z\approx\pm\half$(Anti-Wall). Noting $\pl_\m\Phi=0$, we obtain
\begin{eqnarray}
S=\int d^4x\,dz\frac{2\e^{-4\si(z)}}{F(z)}\sqrt{-g}
\{
-\half (-2F\Si'+5\Si^2)
-\frac{1}{8}{F(z)}^2{\Phi'}^2
-\frac{\la}{4}(\Phi^2-\vz^2)^2-\La\}\nn
\equiv\int d^4x\,dz L\com
\label{concM2}
\end{eqnarray}
where we consider the general curved space $g_\mn(x)$
for the 4D world. \nl
 \nl
i)\ $z\approx 0$
\begin{eqnarray}
F\sim 1\com\q |\Si|\sim \om\com\q
\e^{-4\si}\sim 1\com\q |\Phi|\sim \vz\com\nn
L\sim \sqrt{-g}\{ (2\om M^3-\fourth\vz^2 \frac{L^*}{r_c})\del(z)+\frac{4}{3}\La\}
\pr
\label{concM3}
\end{eqnarray}
The presence of $\del (z)$ term 
\footnote{
In (\ref{concM3}), we have used relations:\ 
$\del (z)=\lim_{a\ra +0}(1/\sqrt{2\pi}a)\exp\{-z^2/2a^2\},\ 
\del(z)\del(z)\sim \del(z)/\sqrt{2\pi}a$, 
where $\sqrt{2\pi}a$ is regarded as the thickness of the wall,
$r_c/L^*$.
}
shows the 3-brane is located at $z=0$.
The brane tension is given by
\begin{eqnarray}
T_{UV}=2\om M^3-\fourth \vz^2 \frac{L^*}{r_c} 
\com
\label{concM3b}
\end{eqnarray}
which is given by the value of $\om=\sqrt{-2\La/3M^3}$ and $\vz$. 
$T_{UV}$ is positive for the large value of $\om$,\ 
$2\om M^3 > \fourth \vz^2 \frac{L^*}{r_c}$.
 \nl
ii)\ $z\approx \half-0$
\begin{eqnarray}
F\sim 2(1-2z)\com\q \Si\sim \om\com\q
\e^{-2\si}\sim (\frac{2}{1-2z})^{-\om/2}\com\q \Phi\sim\vz \com\nn
L\sim \frac{1}{2^\om}\sqrt{-g}\{
(-2\om (1-2z)^\om -\frac{\vz^2}{2}(1-2z)^{\om+1} )\del(z-\half)
+\frac{2}{3}\La (1-2z)^{\om-1}\}\nn
\sim 0\qqqq\qqqq\qqqq\qqqq\com
\label{concM4}
\end{eqnarray}
as far as $\om>1/r_c$. \nl
 \nl
iii)\ $z\approx -\half+0$(or $+\half+0$)
\begin{eqnarray}
F\sim 2(1+2z)\com\q \Si\sim -\om\com\q
\e^{-2\si}\sim (\frac{1+2z}{2})^{+\om/2}\com\q \Phi\sim -\vz\com\nn
L\sim \frac{1}{2^\om}\sqrt{-g}\{
(-2\om (1+2z)^\om -\frac{\vz^2}{2}(1+2z)^{\om+1} )\del(z+\half)
+\frac{2}{3}\La (1+2z)^{\om-1}\}\nn
\sim 0\qqqq\qqqq\qqqq\qqqq\com
\label{concM5}
\end{eqnarray}
as far as $\om>1/r_c$. We see ii) and iii) are continuosly
connected at the Lagrangian {\it density} level. 
Note that singularities at $z=\pm\half$
are avoided by the warp factors. The above result shows
the anti-wall, in the thin wall limit, does not
remain at the lagrangian density level. 
We may say the tension for the
anti-wall is zero.
\begin{eqnarray}
T_{IR}=0
\pr
\label{concM5b}
\end{eqnarray}

This view is also confirmed, 
as far as the value of $\om$ is large, 
by the energy momentum
tensor $\sqrt{-G}T_{MN}$ where $T_{MN}$ is given
by the RHS of eq.(\ref{int2}). 
\begin{eqnarray}
\sqrt{-G}T_\mn=\e^{-6\si}\sqrt{-g}g_\mn
(\fourth F {\Phi'}^2+\frac{2}{F}V)  \nn
\sim
\left\{
\begin{array}{ll}
\sqrt{-g}g_\mn (\fourth\vz^2 \del(z)+2\La) & z\sim 0\ .     \\
\frac{1}{2^{1.5\om}}\sqrt{-g}g_\mn \{
\frac{\vz^2}{2}(1-2z)^{1.5\om+1}\del(z-\half)
+\La (1-2z)^{1.5\om-1}          \}\sim 0
 & z\sim\half-0\ .                                       \\
\frac{1}{2^{1.5\om}}\sqrt{-g}g_\mn \{
\frac{\vz^2}{2}(1+2z)^{1.5\om+1}\del(z+\half)
+\La (1+2z)^{1.5\om-1}          \}\sim 0
 & z\sim -\half+0\ .
\end{array}
\right.                                \nn
\sqrt{-G}T_{zz}=\e^{-4\si}\sqrt{-g}
(-\frac{1}{F}{\Phi'}^2+\frac{8}{F^3}V)  \nn
\sim
\left\{
\begin{array}{ll}
\sqrt{-g} (-\vz^2 \del(z)+8\La) & z\sim 0\ .\\
\frac{1}{2^{\om}}\sqrt{-g} \{
-\frac{\vz^2}{2}(1-2z)^{\om-1}\del(z-\half)
+\La (1-2z)^{\om-3}          \}\sim 0
& z\sim\half-0\ .\\
\frac{1}{2^{\om}}\sqrt{-g} \{
-\frac{\vz^2}{2}(1+2z)^{\om-1}\del(z+\half)
+\La (1+2z)^{\om-3}          \}\sim 0
& z\sim -\half+0\ .
\end{array}
\right.
\nn
\label{concM6}
\end{eqnarray}
(We have assumed $\om>3/r_c$ in the vanishing of $\sqrt{-G}T_{zz}$
for the infrared cases. )

These results are accepted because we have changed
only the infrared boundary condition of the one wall
solution. The appreciable appearance of the anti-wall occurs
only at the level of the equation of motion, not at
the effective action level. 
On the wall the dynamics of the 4D world
are operating, whereas 
it is suppressed (by the warp factor) on the anti-wall. 
From the situation in the 5D lattice simulation,
it is quite interesting to see how
the anti-wall works 
to provide the other chirality partner for the zero mode
fermions bound on the wall. 

\subsection{Renormalizarion Group Flow}
We make a comment from the viewpoint of the renormalization
group in the spirit of AdS/CFT. The basic standpoint is
to regard the 5 dim {\it classical} solution as a scaling trajectory
\cite{Mald98,Witt98,dBVV9912} in the 4 dim {\it quantum} field theory.
In the present case, it occurs near the horizons. As $z\ra (r_c/2)-0$, 
the asymptotic metric (\ref{S.13}) says the scalar density "operator"
of the 5 dim cosmological term is approximated by
\begin{eqnarray}
V(\Phi=\vz)\sqrt{-G}
\approx \La (1-\frac{2z}{r_c})^{\om r_c-1}\sqrt{-g(x)}
\equiv {\bar \La}(z)\sqrt{-g(x)}\com
\label{conc0}
\end{eqnarray}
where the general curved space $g_\mn(x)$ is considered
for the 4D world and ${\bar \La}(z)$ is a 4D scalar. 
We can interpret the meaning of $w_{IR}$ of (\ref{reg4.6})
as the (infrared) scaling parameter when the "scale" $z$
approaches the singular point $r_c/2$.
\begin{eqnarray}
\frac{r_c}{2}-z\sim w_{IR}\approx \frac{1}{8\pi}\frac{r_c}{L^*}
\pr
\label{conc0b}
\end{eqnarray}
From above results, we obtain
\begin{eqnarray}
{\bar \La}(L^*)\approx \La (\frac{1}{4\pi L~*})^{\om r_c-1}\com\nn
\be_{|{\bar \La}|}\equiv \frac{\pl}{\pl (\ln L^*)}\ln~|{\bar \La}(L^*)|
\approx -\om r_c+1=-\sqrt{-\frac{2\La}{3M^3}} r_c+1
\pr
\label{conc0c}
\end{eqnarray}
The last quantity corresponds to the (infrared) renormalization
group ($\be$-) function for $|{\bar \La}|$. We know, from the result
of Sec.6, $-\La$ is positive and is sufficiently large,
hence we conclude $\be_{|{\bar \La}|}<0$. The quantity $|{\bar \La}|$
is infrared asymptotic free, that is, it decreases as $L^*\ra \infty$.
\footnote{
In the renormalizable model of 2 dim $R^2$-gravity, the same
property of the cosmological constant is known\cite{SI95}.
In the brane world context, a similar result is obtained
and analysed in a different treatment\cite{TI0006,Kra0006,Kra0007}.
}

%
%
\section{Discussion and Conclusion}
The regularized solution (\ref{reg4}) cannot become
a true one even when we take $L=\infty$, 
because one of its properties TL5=T5:\ 
symmetric with respect to $z=\fourth$, does not match with
the solution except the $\th$-function limit.
In order to approach a true solution, 
as done for the one-wall case \cite{SI00apr}, we must generalize
the form of solution (\ref{reg4}) 
by replacing the constants $d_\l$ and $c_\l$ by
$z$-dependent functions $d_\l(z)$ and $c_\l(z)$
in the following forms.
\begin{eqnarray}
d_\l(z)=\al_{\l,0}+\frac{\al_{\l,2}}{2!}
\frac{[z^2]}{(L^*r_c)^2}
+\frac{\al_{\l,4}}{4!}
\frac{[z^4]}{(L^*r_c)^4}+\cdots\nn
c_\l(z)=\be_{\l,0}+\frac{\be_{\l,2}}{2!}
\frac{[z^2]}{(L^*r_c)^2}
+\frac{\be_{\l,4}}{4!}
\frac{[z^4]}{(L^*r_c)^4}+\cdots\com
\label{conc1}
\end{eqnarray}
where $[z^n]$ is the "periodic generalization" of
$z^n$ ( in the same way as $F(z)$ in Sec.4).
(Compare with eq.(30) of \cite{SI00apr}.) 
In the above we take only even powers of $z$
in order to keep 
the odd function property ( P4 or TL4=T4).
Note that the above generalization breaks the UV-IR symmetry (T5=TL5).
Therefore the present solution, (\ref{reg4}) with above generalization
(\ref{conc1}),  
can be regarded as the
perturbation around the UV-IR symmetry limit. 

As for the key equations (\ref{reg4}) and their generalization
(\ref{conc1}), we can understand them by a set of general
properties. Let $f(x)$ be a real function defined on 
$x\in \bfR=(-\infty,\infty)$. If $f(x)$ satisfy the following properties:
\begin{description}
\item[G1]
Piecewise continuous everywhere.
\item[G2]
Piecewise smooth everywhere.
\item[G3]
Periodic with the periodicity of $1$
 ($S^1$-symmetry)\ :\ $f(x)=f(x+1)$.
\item[G4]
Odd function ($Z_2$-symmetry)\ :\ $f(-x)=-f(x)$.
(Using the item G3, an important property
:\ $f(\bfZ)=0$ is deduced.)
\item[G5]
Symmetric with respect to $x=\fourth$ (UV-IR symmetry).
\end{description}
then, the general form can be written as 
\begin{eqnarray}
f(x)=\sum_{\l=0}^{\infty}a_\l\sin\{(2\l+1)2\pi x\}
\com
\label{conc2}
\end{eqnarray}
where $\{ a_\l\}$ are constants. If we replace the infinite sum
by the finite sum $\sum_{\l=0}^L$
for the regularization, the word "piecewise" in the items G1 and G2
can always be removed. As a "deformation" of (\ref{conc2}),
at the cost of the item G5 (UV-IR symmetry), we can generalize the
constants $\{ a_\l\}$ to
\begin{eqnarray}
a_\l(x)=\al_{\l,0}+\frac{\al_{\l,2}}{2!}[x^2]
+\frac{\al_{\l,4}}{4!}[x^4]+\cdots\pr
\label{conc3}
\end{eqnarray}
This generalization produces UV$\leftrightarrow$IR 
(Planck$\leftrightarrow$TeV) asymmetry.

Since Ho\v{r}ava-Witten's paper\cite{HW96}, $S^1/Z_2$ compactification
($Z_2$ orbifold)
becomes popular as a dimensional reduction procedure in the string
inspired unified models. It gives essentially an wall at one end
of the extra axis and the anti-wall at the other. 
The present infrared regularization serves as
realizing this configuration.


\vspace{2cm}

\vs 1
\begin{flushleft}
{\bf Acknowledgment}
\end{flushleft}
The author thanks
G.W.Gibbons for stimulating discussions at the initial stage 
and for comments at some stages.
He also thanks T.Tamaribuchi for the help
in the numerical calculation and N.Ikeda for some
discussions.

\vs 1

%
%
\begin{flushleft}
{\Large\bf Appendix A\ :\ Stereographic Compactification}
\end{flushleft}
Instead of the compact coordinate $z$ defined by (\ref{R.2}),
we can take another one $w$ defined as
\begin{eqnarray}
My=\tan \pi\frac{w}{r_c}\com\q
-\half <\frac{w}{r_c}<\half\com\q
-\infty<My<\infty
\pr
\label{sc1}
\end{eqnarray}
See Fig.14. 
\begin{figure}
\centerline{\epsfysize=8cm\epsfbox{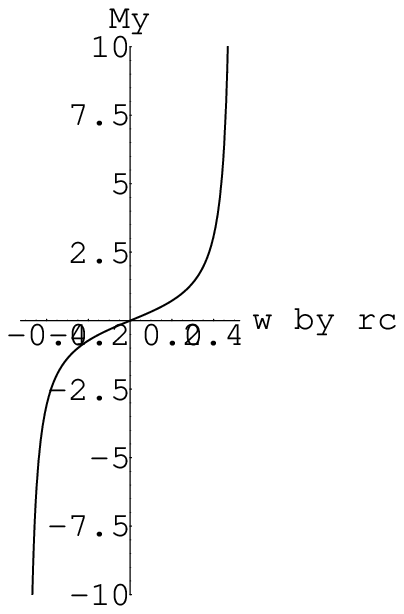}}
   \begin{center}
Fig.14\ Relation between two coordinates,
$w$ (compact) and $y$ (noncompact), (\ref{sc1}).
Vertical axis:\  ${My}$; 
Horizontal axis: $\frac{w}{r_c}$. 
   \end{center}
\end{figure}
We take $M=r_c=1$. The line element (\ref{R.1}) is rewritten as
\begin{eqnarray}
{ds}^2=\e^{-2\si(w)}\eta_\mn dx^\m dx^\n+
\frac{\pi^2}{\{\cos(\pi w)\}^4}{dw}^2\com\nn
dy=\frac{\pi}{\{\cos(\pi w)\}^2}dw
\pr
\label{sc2}
\end{eqnarray}
$w=\pm\half$ are the points of the coordinate singularity.
The boundary condition is
\begin{eqnarray}
\lim_{w\ra\pm(\half -0)}\Phi(w)\ra\pm v_0\com\q
v_0>0\pr
\label{sc3}
\end{eqnarray}
The Einstein equation (\ref{int2}) reduces to the same form as
(\ref{S.5}) but with a different $F$.
\begin{eqnarray}
-\frac{3}{2}(\frac{d\si}{dw})^2{F_1(w)}^2=
-\frac{1}{8}{F_1(w)}^2(\frac{d\Phi}{dw})^2
+V\com\nn
\frac{3}{4}\frac{1}{F_1(w)}\frac{d}{dw}\{\frac{d\si}{dw}F_1(w)\}
=\fourth(\frac{d\Phi}{dw})^2\com\nn
F_1(w)=\frac{2}{\pi} \{\cos(\pi w)\}^2\pr
\label{sc4}
\end{eqnarray}
Compared with the case using
the coordinate $z$, the above equations are
straightforward to the generalization from $w\in (-\half,\half)$
to $w\in\bfR=(-\infty,\infty)$ because $F_1(w)$ is periodic
w.r.t. $w\ra w+1$. 
The properties of $F_1(w)$ is the same as those of $F(z)$
except a slightly better situation in the point F2:\ 
smooth everywhere.
All procedures in the text are valid
for the coordinate $w$ just by replacing $F(z)$ by $F_1(w)$.

In Fig.15, we give a sample result for Vacuum 1w
($\la$=20.0(input), $v_0=1.0$(input), $\La$=-1.888810641,
$\om$=1.122143972). It should be compared with Vacuum 1
in the text. 
The shape of the Higgs potential $V(\Phi)$ is almost
same as that of Vacuum 1(Fig.8), but that of $\Phi(z)$
is much closer to the $\th$-function. 
( In accordance with this, more digits are required for
the appropriate value of the cosmological constant $\La$. )
The "wavy" behavior (App.C)
is recognized in the plot of $\{d_\l\}$. 
The different choice of coordinate gives
the different behaviors such as the sharpness of $\Phi$
(or the value of $w_{UV}$).

\begin{figure}
\centerline{\epsfysize=50mm\epsfbox{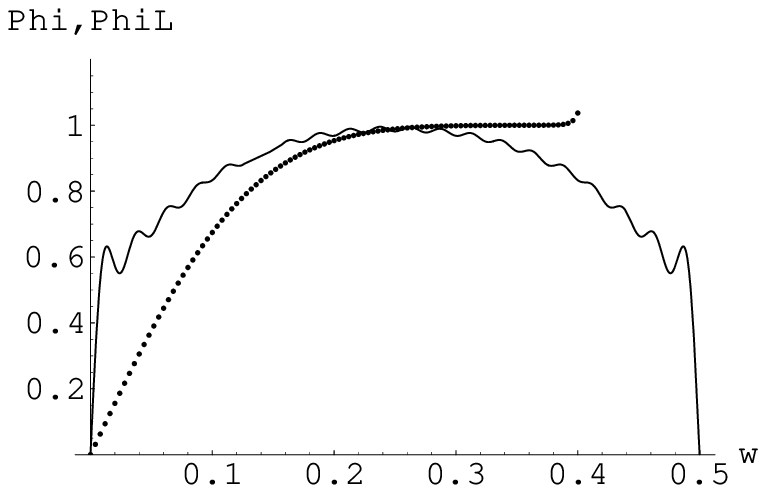}}
\centerline{\epsfysize=50mm\epsfbox{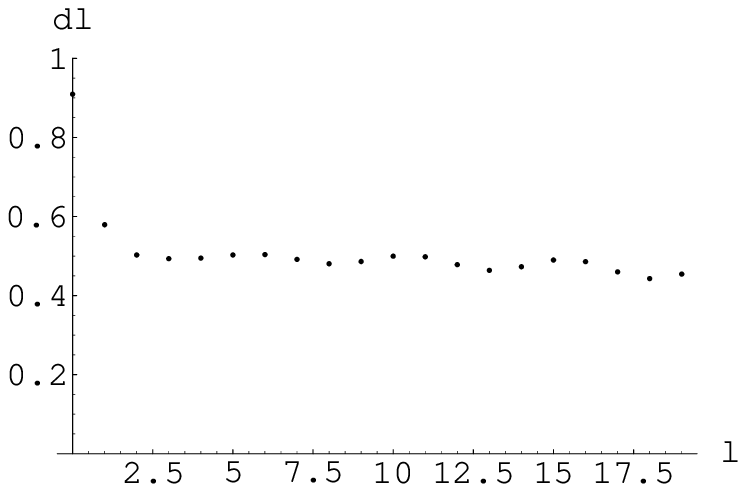}}
   \begin{center}
Fig.15\ Vacuum 1w $(\la,\vz,\La)=(20.0,1.0,-1.888810641)$
:\ [Above] Numerical result of $\Phi(w)$ 
(dotted points, $0\leq w\leq 0.4$)
and its best fit approximate solution $\Phi_L(w)$, (\ref{reg4}).
Horizontal axis is $w$. \newline
[Below] The coefficients \{$d_\l\ ;\l=0,1,\cdots,L=19$\} of $\Phi_L(w)$.
   \end{center}
\end{figure}

\vs 1

%
%
\begin{flushleft}
{\Large\bf
\ Appendix B\ :\ Numerical Results in Sec.6}
\end{flushleft}
The numerical results of Sec.6, where the regularized
solution (\ref{reg4}) of the field equation (\ref{S.5})
is given, are obtained as follow.
\begin{flushleft}
{\large\bf
\ Appendix B.1\ :\ Numerical Solution by Runge-Kutta Method}
\end{flushleft}

First we can directly solve the coupled field equation
(\ref{S.5}) using the numerical method.
In terms of $\Phi(z)$ and $\Si(z)\equiv F(z)\si'(z)$,
the field equation (\ref{S.5}) can be written as
\begin{eqnarray}
\frac{d\Phi}{dz}=\pm\frac{\sqrt{2}}{F(z)}\{
6\Si^2+\la (\Phi^2-{v_0}^2)^2+4\La\}^{\half}\com\nn
\frac{d\Si}{dz}=\frac{2}{3F(z)}\{
6\Si^2+\la (\Phi^2-{v_0}^2)^2+4\La\}\com
\label{rk1}
\end{eqnarray}
where $\la,v_0$ and $\La$ are the vacuum parameters.
$F(z)$ is $(1-4z^2)$ in the text (Sec.4), and 
$(2/\pi)(\cos \pi z)^2$ in App.A. 
Due to the periodicity
and the odd function property, we may focus on
an region $[0,\half)$ of $z$ and
may take only the + sign in the first equation above.  
The above coupled differential equation about
$(\Phi(z),\Si(z))$ can be numerically solved, for a given
vacuum parameters $(\la,v_0,\La)$, by the Runge-Kutta method.
As for the boundary condition we cannot take (\ref{S.8}) and (\ref{S.9}):\ 
$\lim_{z\ra\half -0}\Phi(z)\ra v_0$ and 
$\lim_{z\ra\half -0}\Si(z)= \om=\sqrt{-\frac{2\La}{3}}$,
because of the singularity
at $z=\half$. Instead of this condition in the 
IR (or asymptotic) region, we take
\begin{eqnarray}
\Phi(z=0)=\Si(z=0)=0\pr
\label{rk2}
\end{eqnarray}
This is the condition in the UV (or non-asymptotic) region .
Note that the above one is the {\it necessary condition} 
when we take the boundary condition (\ref{R.5})
or the odd function property of $\Phi(z)$ and $\Si(z)$
(see sentences around (\ref{S.11})). 
We take the following procedure
to have a reliable result.
\begin{enumerate}
\item
Taking the above boundary condition,
we numerically compute (\ref{rk1}) for the region
[0,0.4]. Here we stop the calculation a little before
the singularity point $z=0.5$.
\item
There appear three patterns in the calculational results:\ 
a) They diverge before $z$ reaches $0.4$;\ 
b) They become imaginary before $z$ reaches $0.4$;\ 
c) Calculation lasts to $z=0.4$ and all values
converge to finite ones.
Which pattern appears depends on the choice of the vacuum parameters
($\la,v_0,\La$).
\item
Among the results of pattern c), we pick up the best
one near to the "practical" boundary condition:\  
$\lim_{z\ra 0.4}\Phi(z)\ra v_0$ and 
$\lim_{z\ra 0.4}\Si(z)\ra \om=\sqrt{-2\La/3}$.
Generally, for a given ($\la,v_0$)(input), the calculational
value $\Phi$ or $\Si$ tends to be suppressed (or get imaginary ) when $|\La|$
increases, while it tends to increase ( or diverge)
when $|\La|$ decreases
The width of the wall ( or the initial slope ) can be controlled
by $\la v_0^2$. 
After trying the calculation, with 20-50 values of
$\La$, for one input ($\la,v_0$), we can
always obtain a satisfactory solution. This says that
the boundary condition (\ref{rk2}) and the suitable choice
of $\La$ realize the original boundary condition (\ref{R.5}).
\item
For the numerically unaccessible region (0.4,0.5),
we put the asymptotic values $v_0$ for $\Phi$ 
and $\sqrt{-2\La/3}$ for $\Si$
by hand. Note that this region is the asymptotic region,
therefore it should be dynamically simple
although numerically difficult due to the infrared singularity.
\end{enumerate}
Some sample output data are given in the text ($F(z)=1-4z^2$)
for several cases:\ ($\la,\vz,\La$)\nl
=(20.0,1.0,-1.88855)[Fig.8],(40.0,1.0,-3.77762)[Fig.9],(100.0,1.0,-9.4440537)
[Fig.10,11]. Another one is given in App.A($F(z)=(2/\pi)(\cos\pi z)^2$)
for ($\la,\vz,\La$)=(20.0,1.0,-1.888810641)[Fig.15].

In ref.\cite{SI0107}, using the y-coordinate (\ref{R.1}),
some quantities ($\si'(y), \Phi (y), \Rhat, \mbox{etc}$) are
obtained both analytically and numerically.

\begin{flushleft}
{\large\bf
\ Appendix B.2\ :\ Least Square Fitting}
\end{flushleft}
We fit the solution obtained in the previous subsection
by the proposed 0-th order formula (\ref{reg4}).
$L$ should be taken appropriately large. 
The critical value
$L^*$, explained in App.C, can be roughly obtained by
\begin{eqnarray}
\frac{L^*}{r_c}\sim \sqrt{\la {v_0}^2}
\pr
\label{lsf1}
\end{eqnarray}
Next we fix the coefficients $d's$ and $c's$ of (\ref{reg4}).
A standard way is to minimize the following quantity
(Least square method).
\begin{eqnarray}
I(d_0,d_1,\cdots,d_L)=\int_0^{1/2}(\Phi(z)-\Phi_L(z))^2 dz\com\q
\del I=0\com\nn
J(c_0,c_1,\cdots,c_L)=\int_0^{1/2}(\Si(z)-\Si_L(z))^2 dz\com\q
\del J=0
\com
\label{lsf2}
\end{eqnarray}
where $\Phi(z)$ and $\Si(z)$ are regarded as the exact
solution. Solving this equation, we obtain
\begin{eqnarray}
\frac{v_0}{\pi}\frac{d_\l}{2\l+1}
=\int_0^{1/2}\Phi(z)\sin\{(2\l+1)2\pi z\} dz\com\\
\frac{\om}{\pi}\frac{c_\l}{2\l+1}
=\int_0^{1/2}\Si(z)\sin\{(2\l+1)2\pi z\} dz\com
\pr
\label{lsf3}
\end{eqnarray}
The right hand side of the above equations can be
numerically evaluated using the numerical results of 
$\Phi(z)$ and $\Si(z)$ in the previous
subsection. The samples of $\{d_\l\}$ are given in Fig.8-10 and Fig.15,
and $\{c_\l\}$ are in Fig.11.


%
%
\begin{flushleft}
{\Large\bf Appendix C\ :\ Trapezium Model Solution}
\end{flushleft}

As a simplified model of the solution of (\ref{S.5}) or (\ref{rk1}),
we can take the following simplified model. See Fig.16.
\begin{eqnarray}
\Phi(z)=\left\{ 
\begin{array}{ll}
\frac{\vz}{2w}z  &  \mbox{when}\ 0\leq z\leq 2w \\
\vz              &  \mbox{when}\ 2w<z\leq \half
\end{array}
        \right.
\label{tra1}
\end{eqnarray}

\begin{figure}
\centerline{\epsfysize=50mm\epsfbox{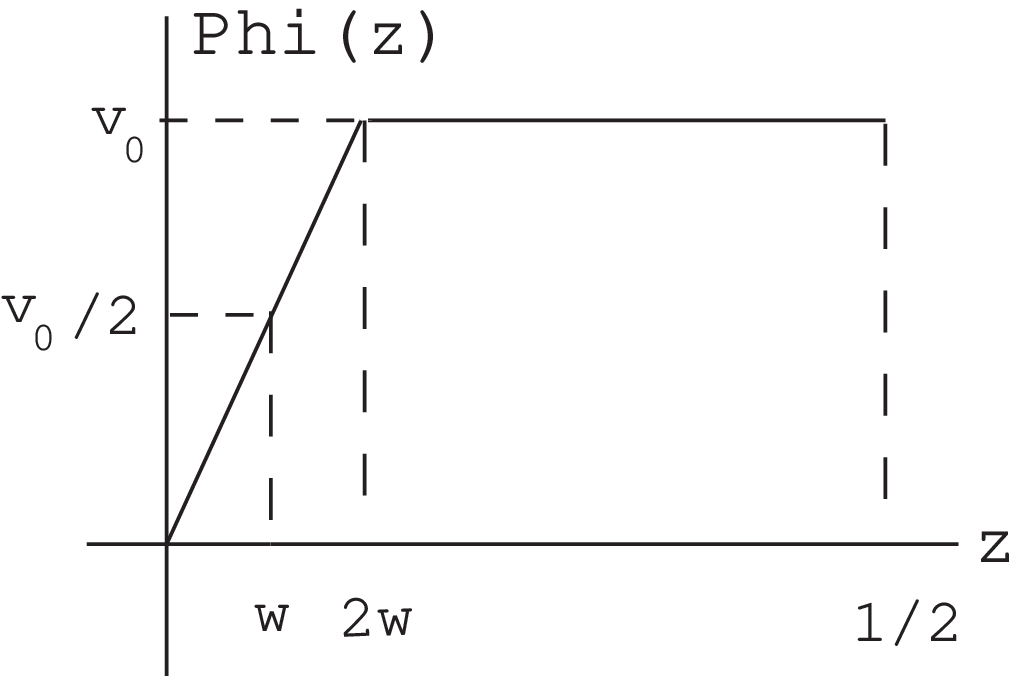}}
   \begin{center}
Fig.16\ The trapezium model solution (\ref{tra1}).
The horizontal axis is $z$, and the vertical one
is $\Phi(z)$.
   \end{center}
\end{figure}

We can qualitatively understand the behavior of the coefficients
\{ $d_\l$ \} appearing in the output data in Sec.6 and App.A. Using
the formula (\ref{lsf3}), $d_\l$ is obtained as
\begin{eqnarray}
d_\l=\half\{
\frac{\sin\{(2\l+1)4\pi w\}}{(2\l+1)4\pi w}+1
\}\com\q \l=0,1,\cdots,L\pr
\label{tra2}
\end{eqnarray}
Generally three "phases" appear depending upon some regions
of $\l$.
\begin{eqnarray}
i)\ 0<(2\l+1)4\pi w\ll 1\q :\q &
d_\l\approx1-\frac{4}{3}\pi^2w^2(2\l+1)^2\ \mbox{(parabolic)}\nn
ii)\ (2\l+1)4\pi w\approx 1\q :\q &
\mbox{wavy region}\nn
iii)\ (2\l+1)4\pi w\gg 1\q :\q &
d_\l\approx \half\ \mbox{(constant)}
\label{tra3}
\end{eqnarray}
The critical value of $\l\equiv L^*$ is given by
\begin{eqnarray}
L^*\approx \frac{1}{8\pi w}\com
\label{tra4}
\end{eqnarray}
which is independent of the regularization parameter $L$.
In Fig.17 and 18, we plot the above result for
($L,w$)=(19,0.15) and ($L,w$)=(19,0.05) respectively.

\begin{figure}
\centerline{\epsfysize=50mm\epsfbox{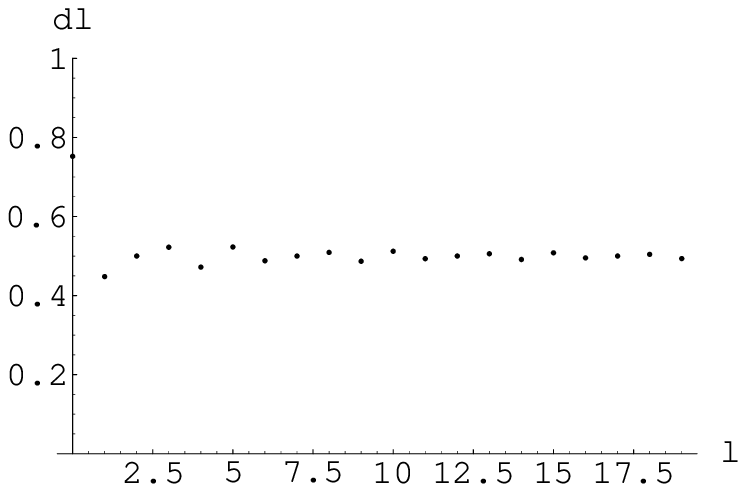}}
   \begin{center}
Fig.17\ The plot of $\{ d_\l\}$ for the trapezium model (\ref{tra2}).
($L,w$)=(19,0.15). 
The horizontal axis :\ $l$; $l$=0,1,$\cdots$,19.
   \end{center}
\end{figure}
\begin{figure}
\centerline{\epsfysize=50mm\epsfbox{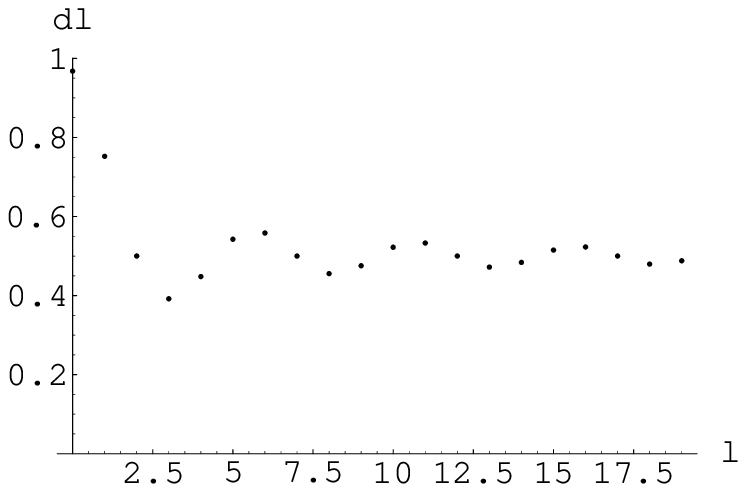}}
   \begin{center}
Fig.18\ The plot of $\{ d_\l\}$ for the trapezium model (\ref{tra2}).
($L,w$)=(19,0.05). 
Axes are the same as in Fig.14.
   \end{center}
\end{figure}



\begin{thebibliography}{99}
\bibitem{RS9905} 
L.Randall and R.Sundrum, {
\PRL {\bf 83}(1999)3370,hep-ph/9905221
}
\bibitem{RS9906} 
L.Randall and R.Sundrum, {
\PRL {\bf 83}(1999)4690,hep-th/9906064
}
\bibitem{Hewett00} 
J.L. Hewett, {Talk at XXXth Int. Conf. on High Energy Physics
(ICHEP2000), Jul.27-Aug.2, 2000, Osaka, Japan,\ 
"Phenomenology of extra dimensions"}
\bibitem{KL0001} 
R. Kallosh and A. Linde,{
JHEP{\bf 0002}(2000)005, hep-th/0001071}
\bibitem{GL0003} 
G.W. Gibbons and N.D. Lambert,{
\PL {\bf B488}(2000)90,
hep-th/0003197.
}
\bibitem{ADD98} 
N.Arkani-Hamed, S.Dimopoulos and G.Dvali,{\PL {\bf B429}(1998)263;
\PR {\bf D59}(1999)086004}
\bibitem{AADD98} 
I.Antoniadis, N.Arkani-Hamed, S.Dimopoulos and G.Dvali,
{\PL {\bf B436}(1998)257}
\bibitem{Mald98} 
J.Maldacena, {Adv.Theor.Math.Phys.{\bf 2}(1998)231,hep-th/9711200}
\bibitem{Witt98} 
E. Witten, {Adv.Theor.Math.Phys.{\bf 2}(1998)253}
\bibitem{dBVV9912} 
J.de Boer,E.Verlinde and H.Verlinde,{
JHEP{\bf 0008}(2000) 003, hep-th/9912012
}
\bibitem{CEHS0001} 
C.Cs\'{a}ki,J.Erlich,T.J.Hollowood and Y.Shirman,{
\NP {\bf B581}(2000)309,
hep-th/0001033.
}
\bibitem{SI00apr}   
S.Ichinose,{Class.Quant.Grav.{\bf 18}(2001)421,hep-th/0003275 
}
\bibitem{ICHEP2000}  
S.Ichinose,{Proc. of the 30th Int. Conf. on High Energy Physics
(ICHEP2000). (Jul.27-Aug.2,2000,Osaka Int. House,Osaka,Japan.)
"An Exact Solution of the Randall-Sundrum Model and the Mass
Hierarchy Problem"}
\bibitem{GW9907b} 
W.D.Goldberger and M.B.Wise,{
\PRL{\bf 83}(1999)4922,hep-ph/9907447
}
\bibitem{Kap92} 
\ D.B.Kaplan,{\PL{\bf B288}(1992)342}
\bibitem{Jan92} 
K.Jansen,{\PL{\bf B288}(1992)348}
\bibitem{CH85} 
C.G.Callan and J.A.Harvey,{\NP{\bf B250}(1985)427}
\bibitem{Sha93} 
Y.Shamir,{\NP{\bf B406}(1993)90}
\bibitem{CH94} 
M.Creutz and I.Horv{\'a}th,{\PR{\bf D50}(1994)2297}
\bibitem{Vra98} 
P.M.Vranas,{\PR{\bf D57}(1998)1415, hep-lat/9705023}
\bibitem{Col0007} 
T.Blum et al,{CU-TP-980,BNL-HET-00/20,RBRC-114,hep-lat/
0007038,"Quenched Lattice QCD with Domain Wall Fermions
and the Chiral Limit"}
\bibitem{Weinb89}  
S.Weinberg,{Rev.Mod.Phys.{\bf 61}(1989)1}
\bibitem{SI84}  
S.Ichinose,{\NP{\bf B231}(1984)335}
\bibitem{GS0005} 
A.Gorsky and K.Selivanov,{\PL {\bf B485}(2000)271,hep-th/0005066}
\bibitem{SI95} 
S.Ichinose,{\NP{\bf B457}(1995)688}
\bibitem{TI0006} 
S.-H.Tye and I.Wasserman, {
\PRL {\bf 86}(2001)1682,
hep-th/0006068. 
}
\bibitem{Kra0006} 
A.Krause,{hep-th/0006226,"A Small Cosmological Constant,Grand Unification
and Warped Geometry"}
\bibitem{Kra0007} 
A.Krause,{hep-th/0007233,"A Small Cosmological Constant and Backreaction
of Non-Finetuned Parameters"}
\bibitem{HW96} 
P.Ho\v{r}ava and E.Witten,
{\NP{\bf B460}(1996)506,hep-th/9510209}
\bibitem{SI0107}   
S.Ichinose,{ 
Univ.of Shizuoka preprint,US-01-03,hep-th/0107254, 
"Some Properties of Domain Wall Solution in the Randall-Sundrum Model"
}
\end{thebibliography}
\end{document}